\documentclass[sigconf]{acmart}

\usepackage{booktabs} 
\setcopyright{rightsretained}
\usepackage{multirow}
\usepackage{amsmath, amssymb}
\usepackage[ruled,vlined,linesnumbered]{algorithm2e}
\usepackage{CJKutf8}
\usepackage[percent]{overpic}
\usepackage{flushend}
\usepackage{ctable}
\newcommand{\cfbox}[2]{%
	\colorlet{currentcolor}{.}%
	{\color{#1}%
		\fbox{\color{currentcolor}#2}}%
}
\def\hlinewd#1{%
	\noalign{\ifnum0=`}\fi\hrule \@height #1 %
	\futurelet\reserved@a\@xhline} 
\newcommand{\prism}{\textit{NamePrism$^e$}~} 
\newcommand{\prismE}{\textit{NamePrism$^e$}} 
\newcommand{\ENEC}{\textit{NamePrism}~}
\newcommand{\ENECW}{\textit{NamePrism$^{w}$}~}
\newcommand{\ENECShort}{\textit{Prism}~}
\newcommand{\ENECWShort}{\textit{Prism$^{w}$}~}
\newcommand{\ENECWE}{\textit{NamePrism$^{w}$}}
\newcommand{\ENECE}{\textit{NamePrism}}

\newcommand*{\affmark}[1][*]{\textsuperscript{#1}}
\newcommand{\fullversion}[1]{}

\begin{document}
\title{Nationality Classification Using Name Embeddings}
\author{%
	Junting Ye\affmark[1], Shuchu Han\affmark[4], Yifan Hu\affmark[2], Baris Coskun\affmark[3]*
	, Meizhu Liu\affmark[2], Hong Qin\affmark[1], Steven Skiena\affmark[1]
}
\affiliation{%
	\institution{$^1$Stony Brook University, $^2$Yahoo! Research, $^3$Amazon AI, $^4$NEC Labs America}
}
\email{{juyye,shhan,qin,skiena}@cs.stonybrook.edu,  {yifanhu, meizhu}@oath.com, barisco@amazon.com}
\thanks{*This research was conducted when Baris Coskun was with Yahoo! Research.}

\begin{abstract}
Nationality identification unlocks important demographic information,
with many applications in biomedical and sociological research.
Existing name-based nationality classifiers use name substrings as features
and are trained on small, unrepresentative sets of labeled names,
typically extracted from Wikipedia.
As a result, these methods achieve limited performance and cannot support
fine-grained classification.

We exploit the phenomena of homophily in communication patterns to learn
\textit{name embeddings}, a new representation that encodes gender,
ethnicity, and nationality which is readily applicable to building classifiers
and other systems.
Through our analysis of 57M contact lists from a major Internet company,
we are able to design a fine-grained nationality classifier covering 39 groups representing over 90\% of the world population.
In an evaluation against other published systems over 13 common classes, our F1 score (0.795) is substantial better than our closest competitor
\textit{Ethnea} (0.580).
To the best of our knowledge, this is the most accurate,
fine-grained nationality classifier available.

As a social media application, we apply our classifiers to the followers of
major Twitter celebrities over six different domains.
We demonstrate stark differences in the ethnicities of the followers of
Trump and Obama, and in the sports and entertainments favored by different groups.
Finally, we identify an anomalous political figure whose presumably
inflated following appears largely incapable of reading the language he posts
in.
\end{abstract}




\maketitle
\section{Introduction}
\label{sec:intro}

Nationality and ethnicity are important demographic categorizations of people, standing in as proxies to represent a range of cultural and historical experiences.
Names are important markers of cultural diversity, and have often served as
the basis of automatic nationality classification for biomedical and sociological research.
For example, nationality from names has been used as a proxy to reflect genetic differences \cite{burchard2003importance,banda2015characterizing} and public health disparity \cite{barr2014health,quesada2011structural} among groups.
Nationality identification is also important in ads targeting, academic studies of political campaigns and social media analysis \cite{chang2010epluribus,appiah2001ethnic}.
Name analysis is often the only practical way to gather ethnicity/nationality annotations, because of privacy concerns.

\begin{figure}[t!]
	\centering
	\includegraphics[width=.5\textwidth]
	{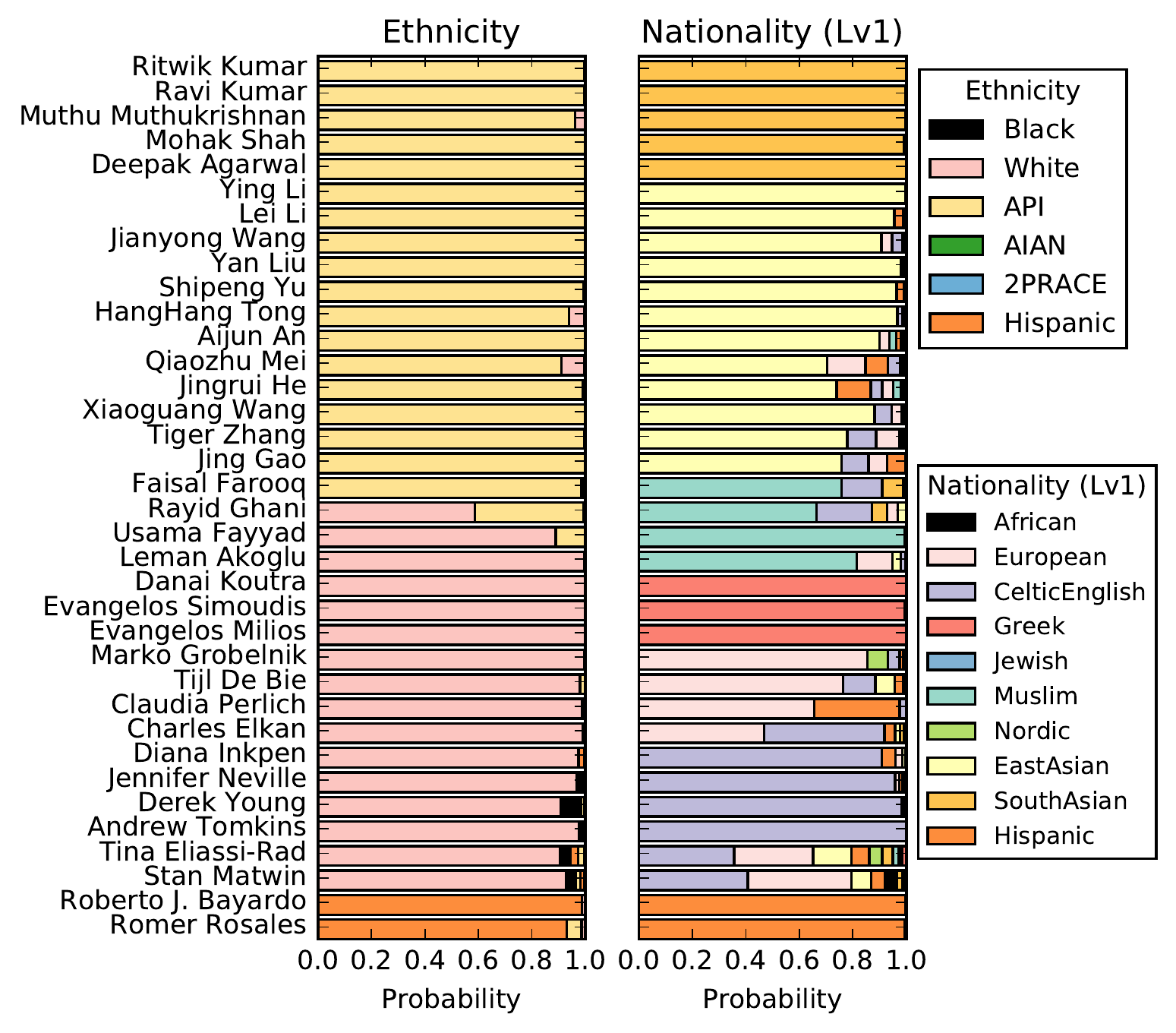}
	\caption{Ethnicity and nationality (Level 1 of the taxonomy) classification on some data mining researchers.}
	\label{fig:kddNames}
\end{figure}

Several previous name-based ethnicity/nationality classification approaches
have been presented \cite{treeratpituk2012name,chang2010epluribus,torvik2016ethnea},
including \cite{ambekar2009name} at KDD '09.
However, the performance of these methods has been constrained by
small and artifical training sets, such as celebrity names from Wikipedia,
and restricted to coarse ethnicity/nationality taxonomies.
The long tail of names makes these approaches 
dependent on surface forms (like substring distributions), which are
by definition ineffective for logograms.
Almost all existing methods are designed only for Latinized names, while other writing systems (e.g. Arabic, Cyrillic) are also widely
used. 

In this paper, we present \ENECE, a new name nationality and ethnicity
classifier which offers a finer-grained taxonomy of ethnic groups.
Fig. \ref{fig:kddNames} demonstrates the performance of our system,
by presenting the ethnicity/nationality probability distributions of
some data mining researchers.
We believe our results will generally agree with the reader's judgement.

Unlike previous methods that rely on substring features,
we propose a more robust representation of names, which exploits the
phenomenon of homophily in communication. The \textit{homophily principle}, that people tend to associate with similar people or
popularly that ``birds of a feather flock together,''
is one of the most striking and empirically robust regularities in social life \cite{mcpherson2001birds, kossinets2009origins}.
Leskovec and Horvitz observed that, in instant messages, people tend to communicate more frequently with others of similar age, language and location \cite{leskovec2008planetary}.
We analyze over 57 million contact lists from an email company, where the account holders are anonymized.
The homophily-induced coherence of these contact lists enables us to derive
meaningful features using
\textit{word embedding} methods \cite{mikolov2013distributed,pennington2014glove} as the basis for a comprehensive and effective nationality classifier.

We collected 74M labeled names come from 118 different countries, containing over 90\% of world's population. We use these labels to define a natural taxonomy of 39 leaf nationalities. As far as we know, our classifier is the most fine-grained and effective one accessible to the public.
The main contributions of our work are:

\begin{itemize}
	\item {\em Introducing Name Embeddings:}
	The contact-list derived name embeddings prove to be a powerful way to capture latent properties of gender, nationality, and age in features readily applicable to classification and regression tasks. Projections of these embeddings are very compelling, creating maps in embedding space that correspond to maps of national boundaries. We believe these embeddings will prove widely applicable to other applications and domains, including those in data privacy and security.
	
	\item {\em Improved Nationality Classification:}
	Our name-based nationality classifier \ENEC performs considerably better than previous classifiers. In particular, on a 13-class evaluation over email/Twitter data, our F1 score (0.795) proves to be much better than competing systems {\em Ethnea\footnote{\url{http://abel.lis.illinois.edu/cgi-bin/ethnea/search.py}}} (0.580) \cite{torvik2016ethnea}, {\em HMM\footnote{\url{http://www.textmap.com/ethnicity/}}} (0.364) \cite{ambekar2009name}, and (on a reduced 10-class scale) {\em EthnicSeer\footnote{\url{http://singularity.ist.psu.edu/ethnicity}}} (0.571) \cite{treeratpituk2012name}. \ENEC uses a Naive Bayes approach within a nationality taxonomy over 39 leaf nodes, employing name embeddings as the primary features.
	
	\item {\em Improved Ethnicity Classification:} A benefit of fine-grained nationality taxonomy is its flexibility to apply to different task settings.The six ethnic groups defined by U.S. Census Bureau over U.S. population largely corresponds to distinct nations of origin. Our ethnicity classifier \prismE, simply reduces the nationality taxonomy from 39 leaf nodes to 6 and incorporates census-based ground truth parameters into the Naive Bayes model.
	
	\item {\em Online Classification Resources:} We release \ENEC as free web service\footnote{\textbf{NamePrism open API}: \url{http://www.name-prism.com/}} for research in sociology, linguistics, and biomedical applications. To the best of our knowledge, it is the only nationality classifier that handles various writing systems, and works on a fine-grained 39-class taxonomy.
	
	\item
	{\em Social Media Analysis:} We use \ENEC to analyze social media, specifically the followers' nationalities/ethnicities of 600 major celebrities on Twitter. Our results show that: (1) Donald Trump's U.S. followers are disproportionally White with followers of Obama and Clinton, (2) ethnicities
	exhibit different preferences in sports and entertainment, and (3) the
	follower counts of a particular Indonesian politician has been artificially inflated by Russian names.
\end{itemize}

The rest of this paper is organized as following. In Sec. \ref{sec:related}, we introduce related works. Sect. \ref{sec:embedding} shows visualization and evaluations of name embeddings. In Sec. \ref{sec:nationality} and \ref{sec:ethnicity}, we describe the methodology and experiments of \ENEC and \prismE. We apply our methods on Twitter celebrities in Sec. \ref{sec:app}. 

\section{Related Work}
\label{sec:related}
Name nationality classification is a fundamental problem with a variety of important applications: \textit{(i)} biomedical research and clinical practice: it is critical to study the genetic and dietary differences among distinct groups\cite{burchard2003importance,banda2015characterizing}. \textit{(ii)} sociology: health care/ employment/ education disparities among different people. \cite{barr2014health,quesada2011structural} \textit{(iii)} online targeting: recommend more accurate ads/news/social media posts to users \cite{chang2010epluribus,appiah2001ethnic}. Other applications includes population demographic studies \cite{aries1989importance,lauderdale2000asian,mateos2007review,mateos2007cultural}. Despite wide-spread demand for nationality labels, it is hard to collect such information via self-reporting because of privacy concerns. Meanwhile, manual annotation of nationality by names is, in fact, a very difficult task, especially for fine-grained taxonomy. 

\begin{figure*}[!ht]
	\begin{minipage}[t]{\textwidth}
		\centering
		\resizebox{1\textwidth}{!}{
			\includegraphics[height=5cm]{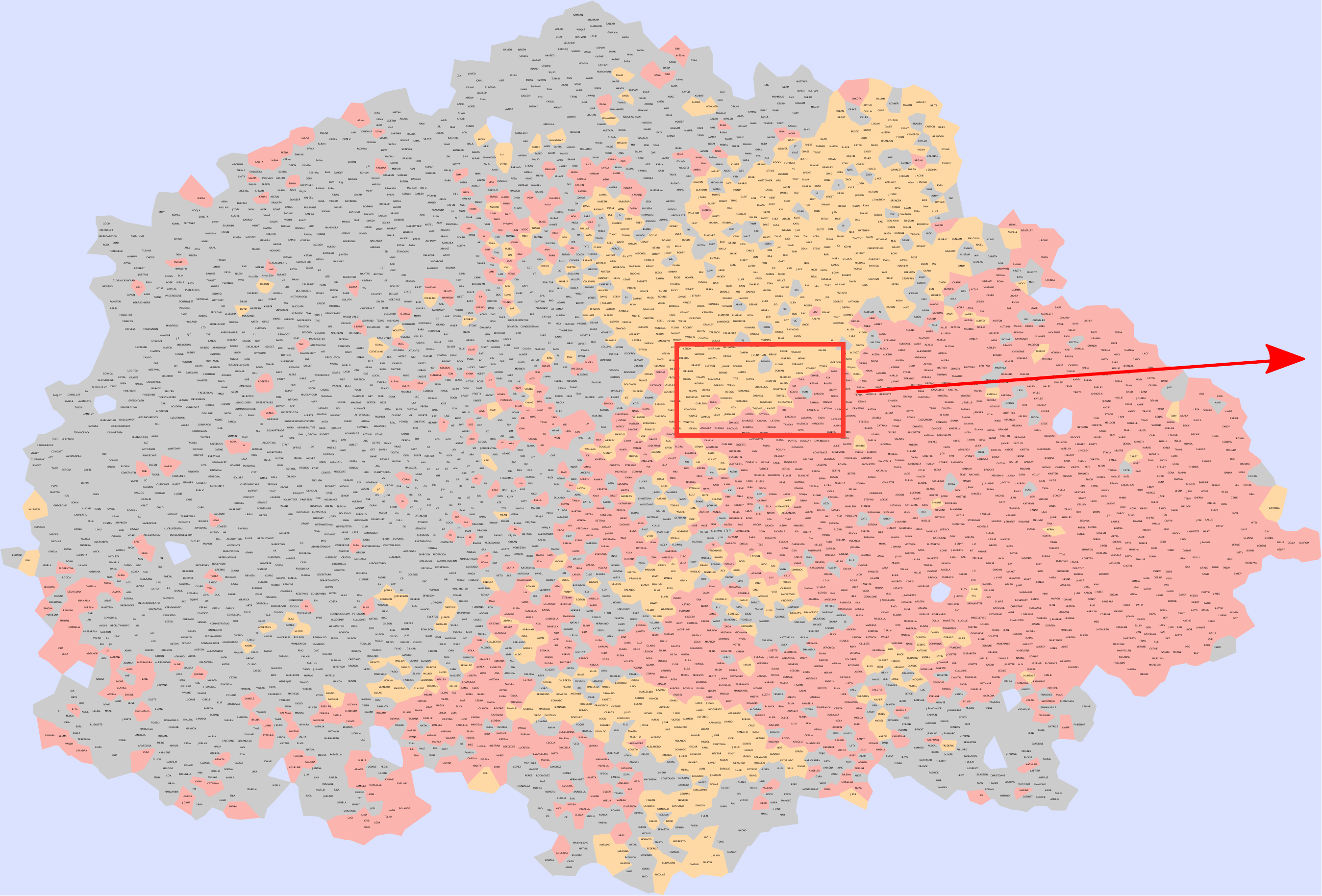}
			\includegraphics[height=5cm]{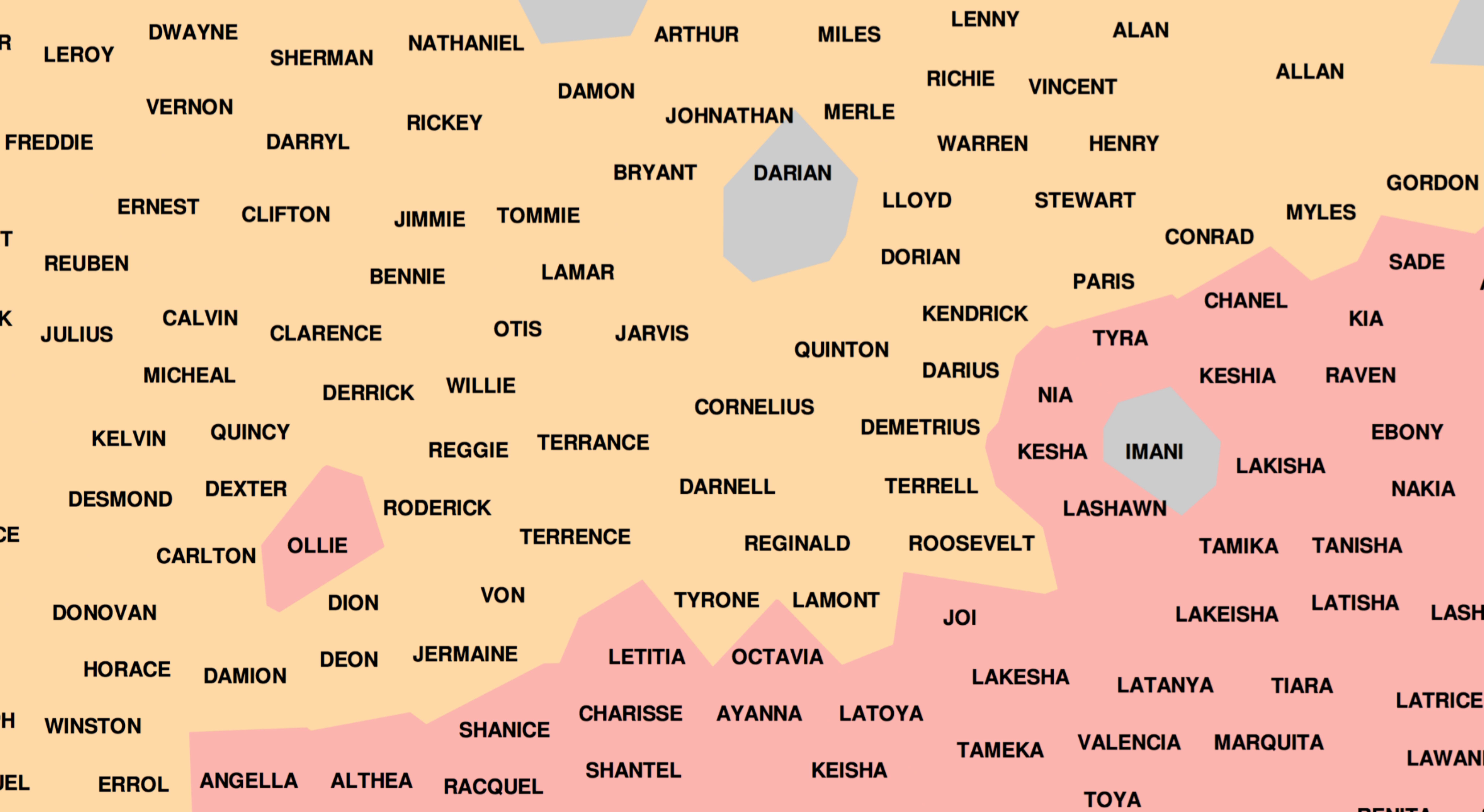}%
		}
		\caption{2D projection (left) of 5K popular first names' \textit{embeddings}. \textbf{Orange} are male names, \textbf{salmon} for females and \textbf{gray} for unlabeled. Same-gender names cluster together, indicating similar embeddings. Inset of the male-female border (right) shows more neutral names. 
			\label{fig:first-name-visualization}}
	\end{minipage}
\end{figure*}

\begin{figure*}
	\begin{minipage}[t]{\textwidth}
		\centering
		\resizebox{1\textwidth}{!}{%
			\includegraphics[height=5cm]{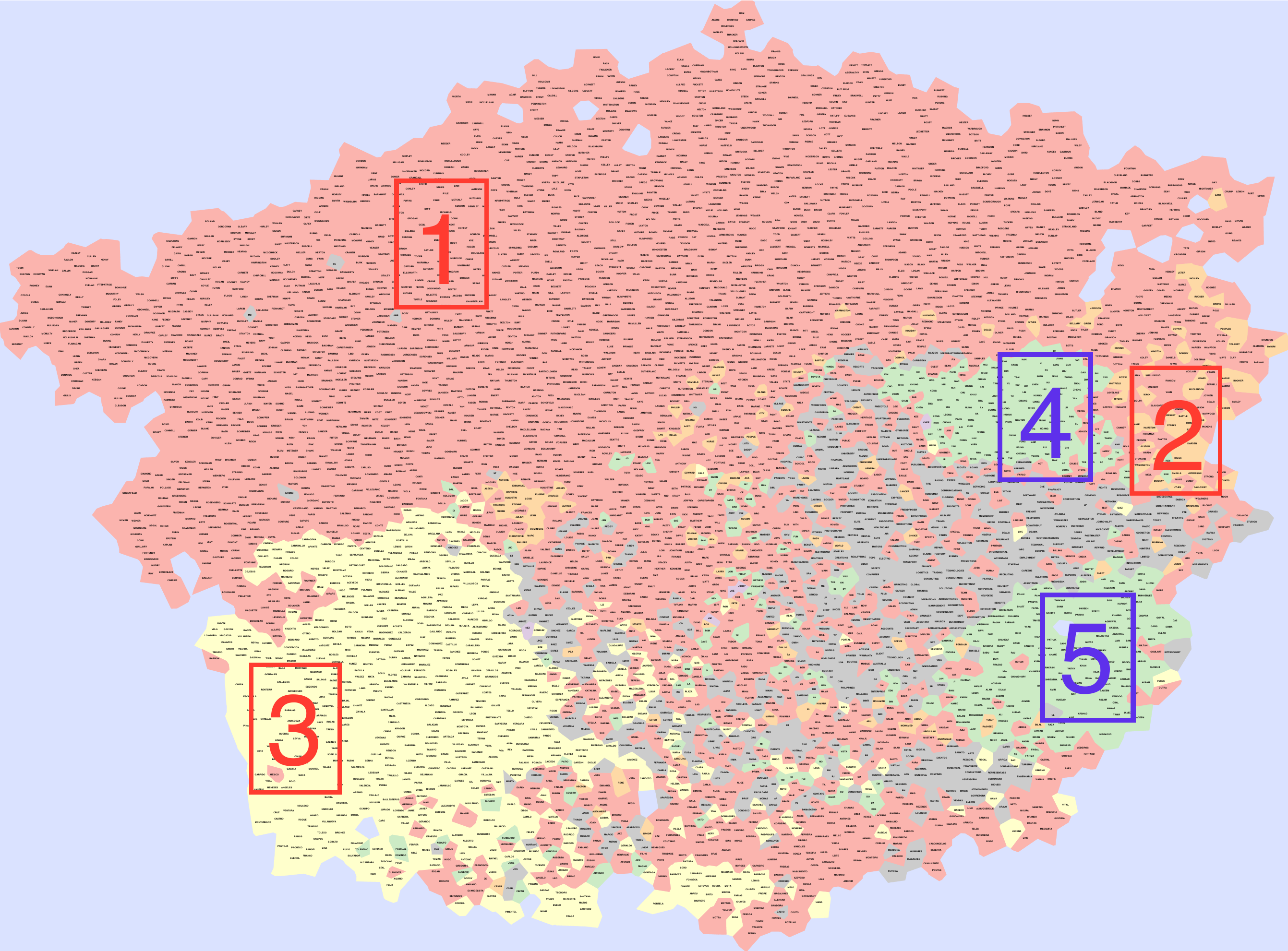}%
			\begin{overpic}[height=5cm]{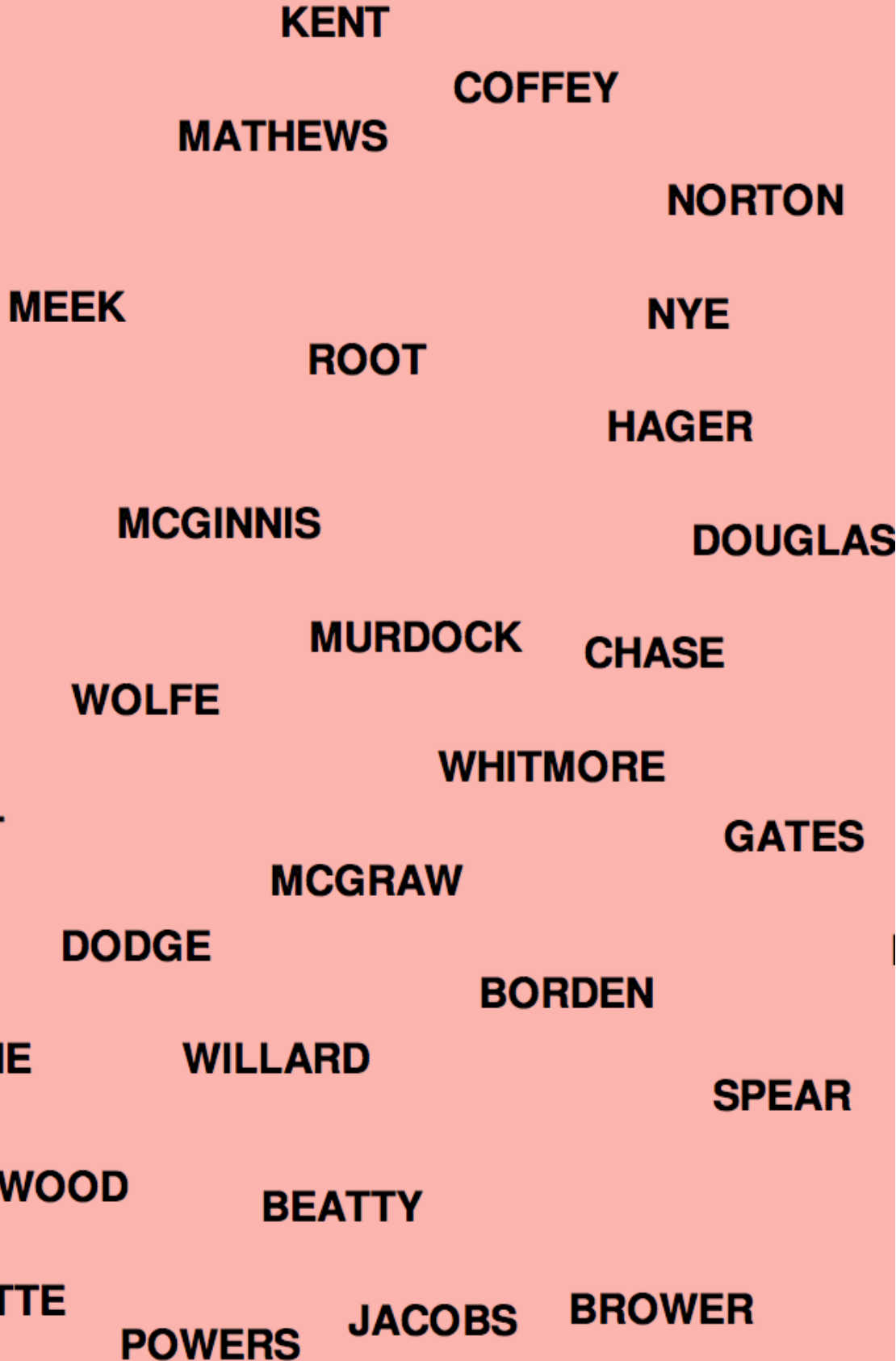}
				\put (28,59) {\Large\cfbox{red}{\textcolor[rgb]{1,0,0}{1}}}
			\end{overpic}
			\begin{overpic}[width=0.192\textwidth]{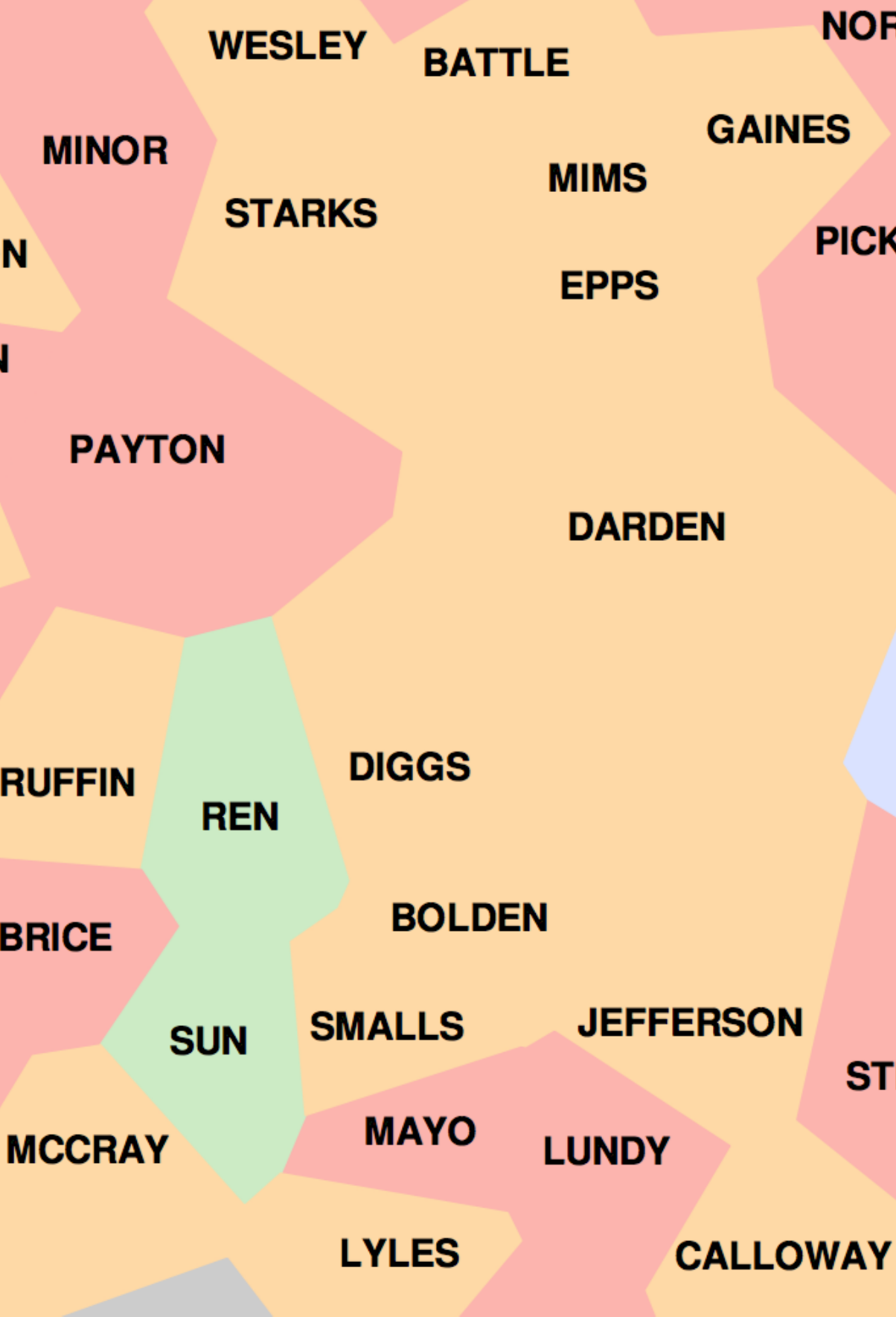}
				\put (45,45) {\Large\cfbox{red}{\textcolor[rgb]{1,0,0}{2}}}
			\end{overpic}
			\begin{overpic}[width=0.203\textwidth]{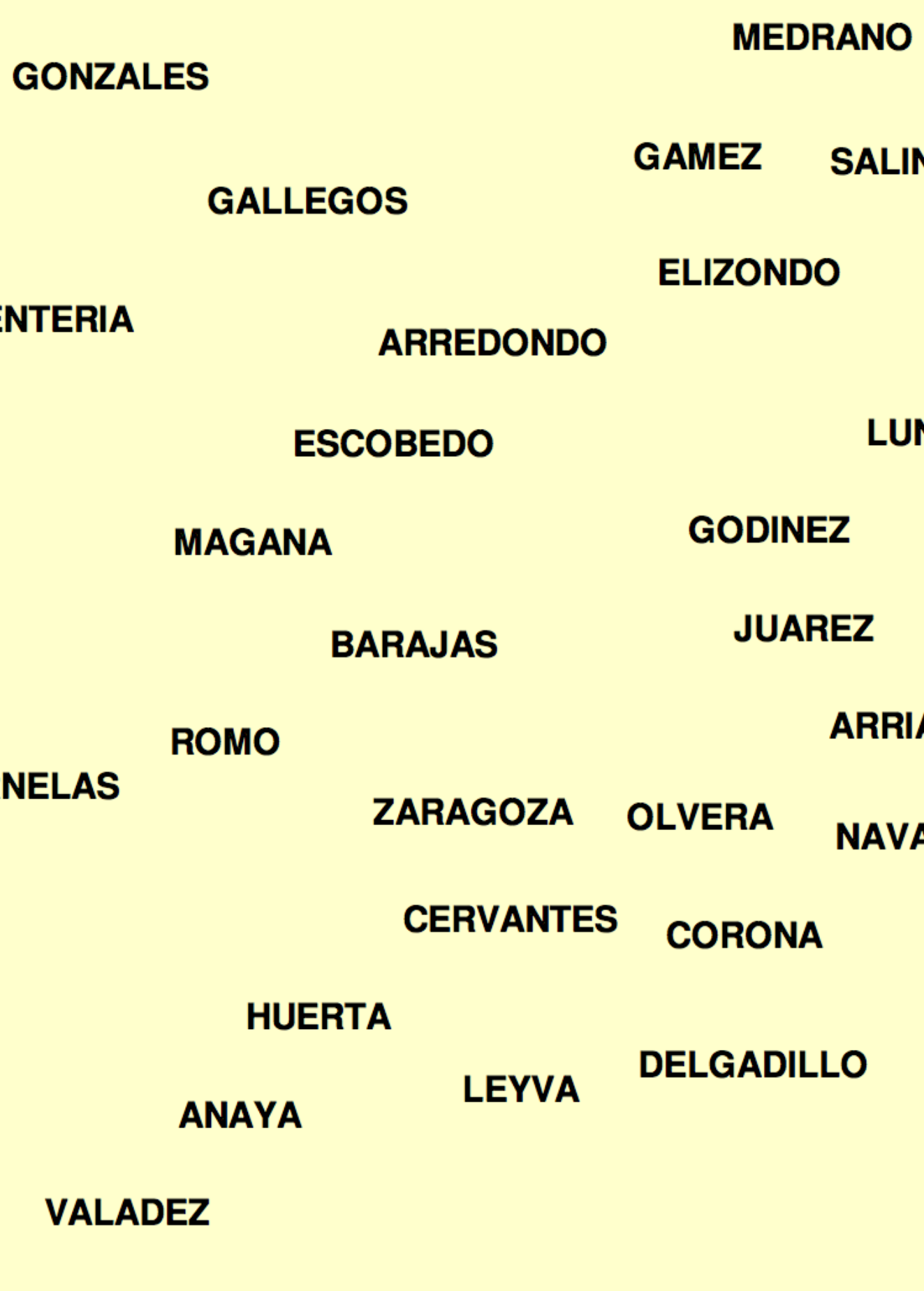}
				\put (41,49) {\Large\cfbox{red}{\textcolor[rgb]{1,0,0}{3}}}
			\end{overpic}
		}
		\caption{2D projection (left) of 5K popular last names' \textit{embeddings}. Same-ethnicity names stand close indicating similar embeddings. Insets (left to right) highlight \textit{White} \fcolorbox{red}{white}{1}, \textit{Black} \fcolorbox{red}{white}{2} and \textit{Hispanic} \fcolorbox{red}{white}{3} names. \textit{API} ( \fcolorbox{blue}{white}{4} and \fcolorbox{blue}{white}{5}) in Fig. \ref{fig:chinese_and_indian}. }
		\label{fig:last-name-visualization}
	\end{minipage}
\end{figure*}

Most recent works use name substrings as features for ethnicity/nationality classification \cite{ambekar2009name, chang2010epluribus, treeratpituk2012name, torvik2016ethnea}. Ambekar et. al. \cite{ambekar2009name} propose to combine decision tree and HMM to conduct classification on a taxonomy with 13 leaf classes. Treeratpituk et. al. \cite{treeratpituk2012name} utilize both alphabet and phonetics sequences in names to improve performance and applied it to analyze how ethnicities evolves in computer science research community \cite{wu2014science}. Chang et. al. \cite{chang2010epluribus} use Bayesian methods to infer ethnicity of Facebook users with U.S. census data and study the interactions between ethnic groups. Torvik and Agarwal \cite{torvik2016ethnea} propose instance-based classifiers by using scientists' names from PubMed. In comparison, we propose name embedding in the light of homophily principle in social life \cite{leskovec2008planetary, kossinets2009origins, mcpherson2001birds}. It is a better representation because substrings are limited to phonogram. Other relevant efforts are binary ethnicity classifiers, including Hispanic \cite{buechley1976generally}, Chinese \cite{coldman1988classification}, South Asian \cite{harding1999potential}.

Name embedding is inspired by word embedding\cite{bengio2003neural,mikolov2013distributed,pennington2014glove}, which has many applications in natural language processing \cite{al2013polyglot,le2014distributed,bengio2014bilbowa}. Other types of data can also benefit from the same assumptions that underlie word embeddings, namely that a data point is governed by the other data in its context \cite{perozzi2014deepwalk, tang2015line,rudolph2016exponential}. \textit{DeepWalk} \cite{perozzi2014deepwalk} learns node embeddings for graph data. It generates contexts by simulating random walks on graphs. Rudolph et. al. \cite{rudolph2016exponential} propose a more general formulation of learning embeddings in different application settings. Similarly, name embeddings treats email contacts with most recency and frequency as context.

\section{Name Embeddings}
\label{sec:embedding}

Name embedding is a variation of word embedding. In a nutshell, word embedding algorithms \cite{bengio2003neural,mikolov2013distributed,pennington2014glove} aim to learn similar embeddings (vectors) if two words co-occur frequently in their contexts. In articles, the context of a word are naturally the words around it. To generate context in contact lists, we need to assign orders to contacts. In the light of homophily principle, we weigh contacts by recency and frequency of communications. As a result, names with large weights tend to have same nationalities. In this way, we construct a ``sentence'' by keeping top contacts of a sorted list. Note that the ordering of sentences in an article is informative for word embeddings. In contrast, the ordering of the contact lists is not useful because email account holders are mutually independent.

\subsection{Name Embedding Visualization}

\begin{figure}[t!]
	\centering
	\begin{overpic}[width=0.223\textwidth]{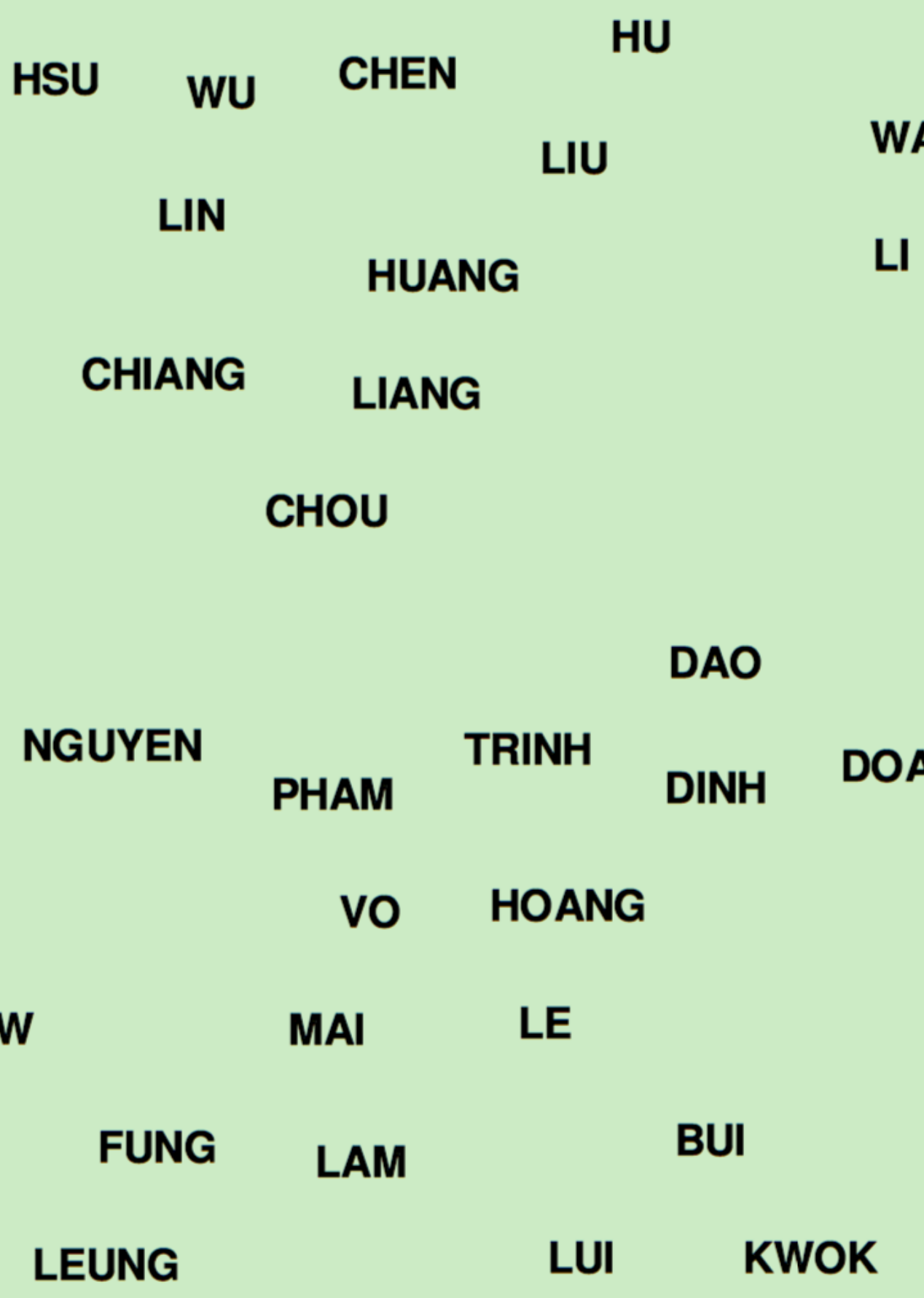}
		\put (40,56) {\Large\cfbox{blue}{\textcolor[rgb]{0,0,1}{4}}}
	\end{overpic}
	\begin{overpic}[width=0.218\textwidth]{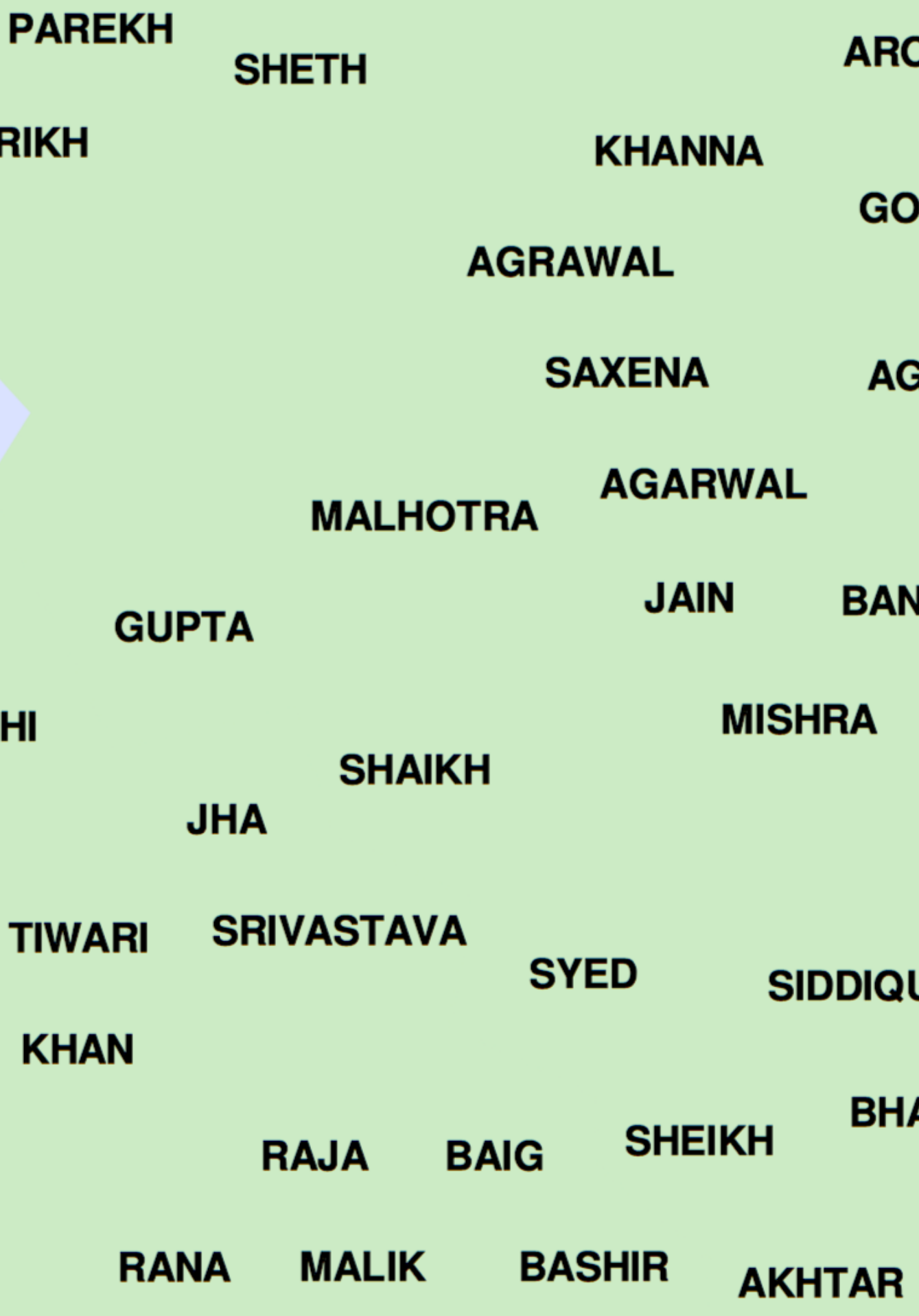}
		\put (18,75) {\Large\cfbox{blue}{\textcolor[rgb]{0,0,1}{5}}}
	\end{overpic}
	\caption{Two distinct Asian clusters. Left: Chinese/ Vietnamese names (\fcolorbox{blue}{white}{4}). Right: Indian names (\fcolorbox{blue}{white}{5}). It shows name embeddings capture nationality signals.} 
	\label{fig:chinese_and_indian}
	\vspace{-0.01in}
\end{figure}

We use t-SNE~\cite{van2014accelerating} to project the 100D name embeddings into 2D, and create the map visualization with {\tt gvmap}~\cite{gansner2010gmap}. U.S. census data are used as ground truths to visualize and evaluate name embeddings. More specifically, we use U.S. 1990 Census data to label popular first names (4.7K female and 1.2K male) and U.S. 2000 Census data to label popular last names (115K White, 5K Black, 6K Asian/Pacific Islander (API), 0.2K American Indian/Alaskan Native (AIAN), 0.1K Two or more races (2PRACE) and 7K Hispanics). As shown from Fig. \ref{fig:first-name-visualization} to Fig. \ref{fig:chinese_and_indian}, genders and ethnicities are labeled with different colors. We are surprised to see how names with same gender, ethnicity and nationality cluster together (Fig. \ref{fig:first-name-visualization} to Fig. \ref{fig:chinese_and_indian}). 

Fig. \ref{fig:first-name-visualization} (left) illustrates the landscape of first names. Using 1990 Census data, we color male names orange, female names pink, and names with unknown gender gray. In general, names of the same gender form mostly contiguous regions. Fig. \ref{fig:first-name-visualization}~(right) is an inset showing a region along the male/female border. We can see that ``Ollie" is labeled as a female name based on Census data (2:1 ratio of female/male instances), while in fact it is often used as a nickname for ``Oliver" or ``Olivia" for daily use. Therefore name embedding is correct in placing it near the border. The embedding also correctly placed ``Imani" and ``Darian", two names not labelled by the Census data, near the border, but in the female/male regions, respectively.

Fig. \ref{fig:last-name-visualization} (left) shows a map of last names.  We color a name according to the dominant ethnicity classification from 2000 Census data. Four major ethnicities are White (pink), Black (orange), Hispanic (yellow), and API (green). Names beyond census data are colored gray. The three insets in Fig. \ref{fig:last-name-visualization} highlight the homogeneity of regions by ethnicities. White, Hispanic and API stand in large contiguous regions while Black are more dispersed. It makes sense because many Black people adopt White names during American slavery time. More interestingly, there are two distinct Asian regions in the map. Fig. \ref{fig:chinese_and_indian} presents insets for these two regions, revealing that one cluster consists of Chinese and Vietnamese names (left) while the other (right) contains Indian names. Even on the left subfigure, Vietnamese names are more gathering around the bottom part while Chinese names on the top. These observations strongly indicate name embeddings capture gender, ethnicity and nationality signals.

\subsection{Evaluation}
We run experiments to validate our observations quantitatively and explore the sensitivity of name embeddings under different parameters. The parameters that we test include: \textit{(i)} different embedding learning method: \textit{CBOW (Continuous Bag Of Word)} or \textit{SG (Skip-Gram)}; \textit{(ii)} use joint embedding space of first/last names or separate; \textit{(iii)} number of nearest neighbor.

We can see from Tab. \ref{embedding-comparison-table} that the joint variants generally perform best. However the differences between the variants are relatively small. In addition, the \textit{CBOW} model generally outperforms the \textit{SG} model. It seems $P_1(B|B)$ is relatively low (0.35-0.59). However, it is essentially a harder task to find a black name because a random name from the contact lists has a probability of 0.03 being Black, while 0.74 being White.

\section{Nationality Classification}
\label{sec:nationality}
\subsection{Methodology}
\ENEC uses Naive Bayes model because of its effectiveness and interpretability. We argue that name nationalities depend on both first name and last name. This is especially effective for names used across different nationalities but with different popularities. It also helps to reduce errors when names are mixtures because of immigration or cross-nationality marriages. We put much effort on estimating parameters, i.e. name parts likelihood, using features from training data, name embedding, substrings and string characters. Therefore, each parameter has at most 4 estimations. NamePrism uses the ones with largest confidence for predictions.

\begin{table}[t!]
	\small
	\centering
	\begin{tabular}{@{}|c|c|cc|cc|@{}} \hline
		\multirow{2}{*}{} &  \multirow{2}{*}{Metrics} & \multicolumn{2}{c|}{Joint} &   \multicolumn{2}{c|}{Seperate}  \\  \cline{3-6}
		&  & CBOW & SG & CBOW & SG \\ \hline
		\multirow{2}{*}{Gender} & $P_1(G_i|G_i)$ & 0.909 & 0.884 & \boldmath\textbf{0.916} & 0.884\\ 
		& $P_{10}(G_i|G_i)$ & \boldmath\textbf{0.936} & 0.927 & 0.935 & 0.921\\ \hline
		\multirow{4}{*}{Ethnicity} & $P_1(W|W)$ & 0.936 & \boldmath\textbf{0.946} & 0.930 & 0.922\\ 
		& $P_1(B|B)$  & \boldmath\textbf{0.594} & 0.456 & 0.444 & 0.345\\ 
		& $P_1(A|A)$ & \boldmath\textbf{0.763} & 0.721 & 0.717 & 0.680\\ 
		& $P_1(H|H)$ & \boldmath\textbf{0.754} & \boldmath\textbf{0.754} & 0.671 & 0.697\\ \hline
	\end{tabular}
	\caption{Evaluations of different name embedding variants. CBOW and SG are two word embedding methods. $P_1(G_i|G_i)$ is the probability that 1 nearest neighbor (1-NN) is of the same gender while $P_{10}(G_i|G_i)$ is for 10-NN. ``W'', ``B'', ``A'', ``H'' stand for ``White'', ``Black'', ``API'', ``Hispanic'', respectively.}
	\label{embedding-comparison-table}
	\vspace{-0.2in}
\end{table}

\subsubsection{Naive Bayes Model}
\label{subsubsec:NB}
In many case, our last names reveal our nationality origins. For example, ``Zhang'' is a common Chinese last name. It is easy to predict one's nationality if his last name is unique to that nation. However, there are many last names that are popular across nationalities. For example, ``Lee'' is popular in both China (especially in Hong Kong) and the UK. For ``Qiang Lee'' and ``John Lee'', we would make mistakes if we only take signals from the last name. Combining with first names, we can perform better because it is easy to see whether the first name is more in China or UK. Similarly, using both name parts also helps when names are mixtures due to immigration or cross-nationality marriage.

Our method, \ENECE, can be formalized in Eq. \ref{equ:NB}: 

\begin{equation}
\label{equ:NB}
P(N|v_f,v_l) \propto P(v_f|N)  P(v_l|N)  P(N)
\end{equation}

\noindent where $N$ denotes nationality, $v_l$ means last name and $v_f$ is first name. We will describe our methods to estimate the likelihood (i.e. $P(v_f|N)$, $P(v_l|N)$) for frequent and rare names in next subsection. We can get Equ. \ref{equ:NB} by using Bayesian rule under the assumption that $v_f$ and $v_l$ are conditionally independent given $N$.

\subsubsection{Parameter Estimation}
\label{subsubsec:paraEstimation}
We estimate name part likelihood from 4 sources: \textit{(i)} \textit{training data}, i.e. the names appear in training data (denoted as $V_{tr}$); \textit{(ii)} \textit{name embedding}, the names from contact lists that have embeddings ($V_{em}$); \textit{(iii)} \textit{prefix/suffix strings}, names that share the same prefix/suffix with names in training data ($V_{p/s}$); \textit{(iv)} \textit{name characters}, names that use the same language characters (e.g. Arabic) seen in training data ($V_{ch}$). Intuitively, the increasing order of vocabulary size is $V_{tr}$, $V_{em}$, $V_{p/s}$, $V_{ch}$, which is also the decreasing order of estimation confidence. 

\paragraph{Training Data} 
Eq. \ref{equ:traindata} shows the most effective and simple way to estimate $P(v_l|N)$ and $P(v_f|N)$ directly from training data. 

\begin{equation}
\label{equ:traindata}
P_{tr}(v_i|N) = \frac{C(v_i, N)}{C(N)} , v_i \in V_{tr}
\end{equation}

\noindent where $v_i$ is either a first name or last name from $V_{tr}$. $C(v_i, N)$ is the count of $v_i$ with nationality $N$ and $C(N)$ is equivalent to $\sum_{v} C(v,N)$. Note that each name part in $V_{tr}$ have more than 5 occurrences in training data so that we have high confidence in the estimation.

\paragraph{Name Embedding}
The likelihood of names ($v_i$) in $V_{em}$ can be estimated using k-NN, i.e. take the average of k nearest neighbors (e.g. kNNs) in $V_{tr}$. However, we did not directly estimate the likelihood using its kNNs' likelihood. Instead, we realize that it performs better if we first estimate $v_i$'s posterior using its neighbors' posteriors and then apply Bayes rule to estimate the likelihood (Eq. \ref{equ:embdNB}). It makes sense because names with similar embeddings do not necessarily have similar popularity (see Fig. \ref{fig:last-name-visualization}). The estimation of $P_{em}(v_i|N)$ is formulated by Eq. \ref{equ:embdNB} and \ref{equ:embdEsti}.

\begin{equation}
\label{equ:embdNB}
P_{em}(v_i|N) = \frac{P_{em}(N|v_i)  P(v_i)}{P(N)}, v_i \in V_{em}
\end{equation}

\begin{equation}
\label{equ:embdEsti}
P_{em}(N|v_i) = \frac{1}{|kNN(v_i)|}\sum\limits_{v_j\in kNN(v_i)} P_{tr}(N|v_j)
\end{equation}

\noindent where $kNN(v_i)$ is the set of name parts that are $v_i$' kNNs.

\paragraph{Prefix/Suffix Strings}
As mentioned in \cite{ambekar2009name}, prefix and suffix of name parts are indicative features. For name part $v_i \in V_{p/s}$, we can estimate its likelihood by averaging the ones' which share the same prefix/suffix.

\begin{equation}
\label{equ:preSuf}
P_{p/s}(v_i | N) = \frac{1}{|PS(v_i)|}\sum\limits_{v_j \in PS(v_i)}P_{tr}(v_j | N), v_i \in V_{p/s}
\end{equation}

\noindent where $PS(v_i)$ is the set of prefix and suffix strings of $v_i$. Here we use substrings with length between 3 to 5. $P_{p/s}(v_j | N)$ is the average likelihood of name parts in $V_{tr}$ that have prefix/suffix $v_j$.

\paragraph{Name Characters}
\label{subsubsec:NC}
If a name is so rare that it is not in $V_{tr}$ nor $V_{em}$. Moreover, it doesn't contain valid prefix or suffix strings. For example, a name written in ``Hangul'',
\begin{CJK}{UTF8}{mj}
	``근혜''
\end{CJK}
. It is very likely to be a Korean name because most names in ``Hangul'' are Korean names. Therefore, for a name $v_i \in V_{ch}$, we use the average of names in same characters to estimate its likelihood.

\begin{equation}
\label{equ:character}
P_{ch}(v_i | N) = \frac{1}{|CH(v_i)|}\sum\limits_{v_j \in CH(v_i)}P_{tr}(v_j | N), v_i \in V_{ch}
\end{equation}

\noindent where $CH(v_i)$ is the set of names in training data that are written in the same language as $v_i$.

\begin{algorithm}[t!]
    \SetKwInOut{Input}{Input}
    \SetKwInOut{Output}{Output}
    \Input{first/last name $v_f$, $v_l$; nationality taxonomy; estimated parameter sets $P_{tr}, P_{em}, P_{p/s}, P_{ch}$.}
    \Output{nationality prediction $T$}
    \textbf{Init} $T = $ root class \;
    \While{$T$ is not a leaf class}{
        \For{child class $N_i$ of \hspace{0.05in}$T$}{ 
            \For{each name part $v\in\{v_f, v_l\}$}{
                \uIf{$v \in V_{tr}$}{ 
                    $P(v|N_i) = P_{tr}(v|N_i)$; 
                }
                \uElseIf{$v \in V_{em}$}{
                    $P(v|N_i) = P_{em}(v|N_i)$;
                }
                \Else{
                    $P(v|N_i) = \sigma$; \hfill \# $\sigma$ is a small constant\
                }
            }
        }
        \If{ neither of $v_f, v_l$ in $V_{tr}$ or $V_{em}$}{
            \For{child class $N_i$ of \hspace{0.05in}$T$}{ 
                \For{each name part $v\in\{v_f, v_l\}$}{
                    \uIf{$v \in V_{p/s}$}{ 
                        $P(v|N_i) = P_{p/s}(v|N_i)$; 
                    }
                    \uElseIf{$v \in V_{ch}$}{
                        $P(v|N_i) = P_{ch}(v|N_i)$; 
                    }
                }
            }
        }
        $P(N_i | v_f, v_l) \propto P(v_f|N_i) \cdot P(v_l|N_i) \cdot P(N_i)$\;
        $T = \arg\max\limits_{N_i}P(N_i | v_f, v_l)$\;
    }
    \textbf{return} $T$\;
    \caption{\ENEC, a hierarchical nationality classifier}
    \label{alg:hierarchical}
\end{algorithm}

\subsubsection{Internet Population vs. World Population}

\label{subsubsec:worldPop}
As we have shown in previous subsections, the name parts likelihood are estimated from email/Twitter users. However, Internet services (Email and Twitter) has varying popularity in different countries. Therefore, we need to assign different priors if a name is not sampled from Internet users. For example, UK and South Africa have similar population (around 50M to 60M). In our datasets, we have an order of magnitudes more names from the UK than from South Africa. Therefore we need to adjust to the real population of countries when we are predicting a random name from the world population. 

\begin{figure*}[!ht]
	\centering
	\includegraphics[width=1\textwidth]%
	{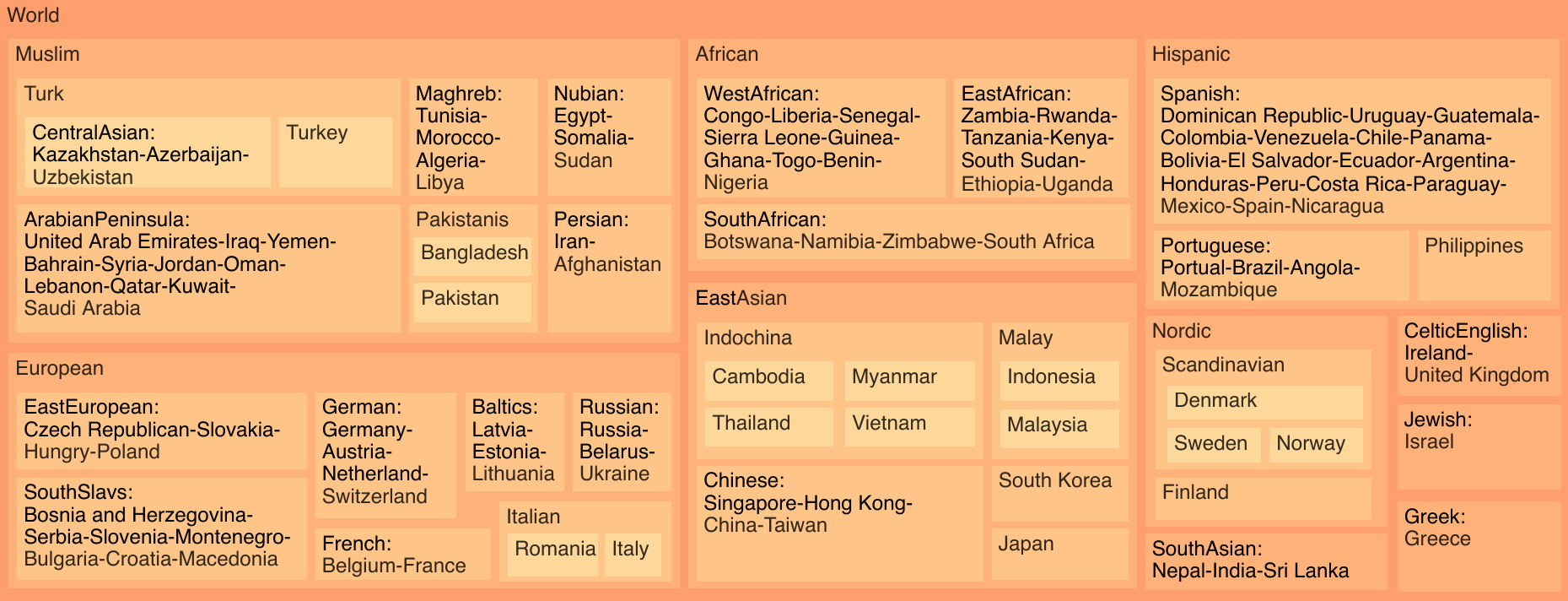}
	\caption{Treemap of nationality taxonomy. Nested blocks within a larger block are its child nodes. 118 countries/regions, covering over 90\% world population, are assigned to 39 leaf nationalities. The taxonomy is constructed based on Cultural, Ethnic and Linguist (CEL) similarities. 
	}
	\label{fig:classTree}
\end{figure*}
\begin{figure}[!ht]
	\centering
	\includegraphics[width=0.4\textwidth]%
	{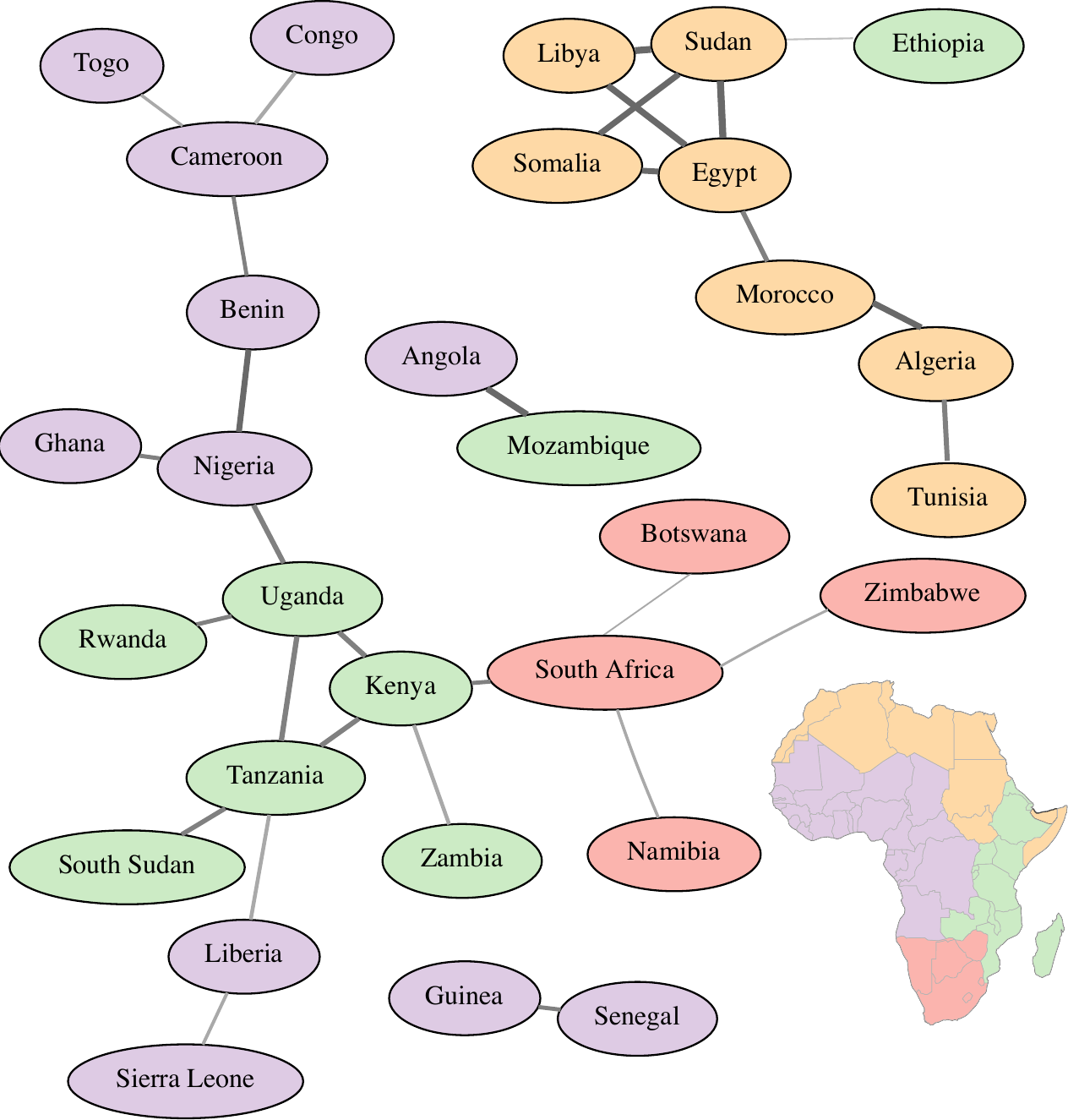}
	\caption{Name similarities between countries of Africa using Email/Twitter data. Thicker edges indicate stronger similarities and more common first/last names usages. Eastern African countries in green, western in purple, northern in orange and southern in red. The clusters of same-color nodes indicate that countries with similar names tend to be close geographically.}
	\label{fig:sim_Africa}
\end{figure}

Formally, let  $P^I(\cdot)$ be probabilities over Internet population, and $P^W(\cdot)$ be the probability over world population. We have $P^I(v_i|N) = P^W(v_i|N)$ by assuming that names of Internet population are random samples from corresponding countries. Let $I^N$ be the number of names with nationality $N$ on Internet population and $W^N$ be the one of $N$ on world population. Thus, $P^I(N) = \frac{I^N}{\sum_N I^N}$ and $P^W(N) = \frac{W^N}{\sum_N W^N}$. We can get the relation between $P^I(N)$ and $P^W(N)$ with Eq. \ref{equ:YahooVsWorld}.

\begin{equation}
\label{equ:YahooVsWorld}
P^I(N) = \frac{I^N}{\sum_N I^N} = \frac{W^N}{\sum_N W^N}  \frac{\frac{I^N}{W^N}}{\frac{\sum_N I^N}{\sum_N W^N}} = P^W(N)  \frac{s_N}{S}
\end{equation}

\noindent where $S$ is the overall sample ratio and $s_N$ is the sample ratio of $N$. $P^I(N)$ can be estimated from training data. $s_N$ and $S$ can be computed by looking up countries' populations. We can put Eq. \ref{equ:YahooVsWorld} into Eq. \ref{equ:NB} when classifying names from world population.

\subsubsection{Hierarchical Classification}
Names are classified on a predefined taxonomy in top-down fashion (see Fig. \ref{fig:classTree}). The detailed algorithm are shown in Alg. \ref{alg:hierarchical}. We start from root class of the taxonomy (line 1). In each iteration, it picks the class that maximizes $P(N_i | v_f, v_l)$ (from line 2 to 19) until it meets a leaf class. Since we have higher confidence in $P_{tr}$ than $P_{em}$, so we prefer parameters from the former (line 3 to 10). If neither of the name parts are in $P_{tr}$ or $P_{em}$, we use the parameters from $P_{p/s}$ or $P_{ch}$ (line 11 to 17). Note that if only one of the name part in $P_{tr}$ or $P_{em}$, we will only use the partial signal and smooth the other name part.

\subsection{Nationality Taxonomy Construction}
The nationality taxonomy is a key component in our method. Mateos et al. proposed a nationality taxonomy based on Cultural, Ethnic and Linguist (CEL) similarities \cite{mateos2007cultural}. Our name-based nationality taxonomy is constructed on top of CEL-based taxonomy, especially for the top level construction. While there is no ``gold standard'' name-based nationality taxonomy because of the complexity in naming customs around the world, we consult opinions from linguists and people from different cultures to reach a common ground as a useful approximation . Moreover, as shown in Sec. \ref{subsubsec:countrySim}, we could compute similarities between countries using name parts distributions. These similarities are helpful to construct the bottom levels of the taxonomy. For example, Hispanic countries are divided into three subgroups: Spanish, Portuguese and Philippines. The reason is, countries within Spanish and Portuguese are very similar to each other according to name similarities, indicting finer-granularity groupings are not feasible and necessary.  

\subsection{Datasets}
\subsubsection{Name Labels}
In order to estimate parameters mentioned above, we need name labels, i.e. full name and nationality pairs. We collected 68M such pairs from the Email source and 6M pairs from Twitter, totaling 74M labeled names from 118 major countries (Fig. \ref{fig:classTree}). These countries take up over $90\%$ of world population. To remove noise, we filter out names where both parts appear only once. 91\% names remain. Note that we are interested in nationalities, thus immigration countries, including U.S., Canada and Australia, are not included in our dataset. To preserve privacy for the email data, the IDs of these users (e.g. email address) are removed. Furthermore, we only retain the counts of first/last names and countries. We used the full name and country labels solely for the purpose of performance measurement. They are not retained for classification.

\paragraph{Email}
90\% of name labels come from email. Each name part appears at least twice so that typos and random strings are filtered. Note that the email contact lists and labeled names are different set of users. We set 5 as thresholds for both $V_{tr}$ and $V_{em}$. It turns out $|V_{tr}|$ is 1.02M and $|V_{em}|$ is 4.09M. It makes sense because contact lists are names from many email companies and thus a larger population.

\begin{table*}[!ht]
	\centering
	\begin{tabular}{@{}|c|rccccc|rccccc|@{}} \hline
		&  \multicolumn{6}{c|}{Wikipedia Data} &  \multicolumn{6}{c|}{Email/Twitter Data} \\ \hline
		Nationality 			& Name\#& \textit{HMM}   & \textit{Ethnea} & \textit{Embd}  & \ENECShort &\ENECWShort 		&Name\# & \textit{HMM}   &\textit{Ethnea} & \textit{Embd}  & \ENECShort & \ENECWShort \\ \hline
		GreaterAfrican 		& 11K 	& 0.428	& 0.532 & 0.480 & \textbf{0.543}& 0.486	& 31K	& 0.269 & 0.389 & 0.554 &\textbf{0.645} & 0.622\\
		GreaterEuropean 	& 113K 	& 0.863 & 0.903 & 0.927 &\textbf{0.932} & 0.899 & 225K 	& 0.725 & 0.815 & 0.861 &\textbf{0.920} & 0.902\\ 
		Asian 				& 24K 	& 0.654 & 0.670 & 0.711 & 0.745 &\textbf{0.748} & 123K 	& 0.674 & 0.709 & 0.763 &\textbf{0.910} & 0.904\\ \specialrule{1.5pt}{1pt}{1pt}
		Muslim*				& 7K 	& 0.380 & 0.563 & 0.538 &\textbf{0.615} & 0.611 & 13K 	& 0.204 & 0.374 & 0.602 &\textbf{0.612} & 0.533\\ 
		Africans* 			& 4K 	& 0.285 & 0.268 & 0.282 &\textbf{0.314} & 0.259 & 18K 	& 0.174 & 0.288 & 0.458 & 0.636 &\textbf{0.659}\\ 
		WestEuropean 		& 49K 	& 0.631 & 0.724 & 0.709 & 0.747 &\textbf{0.756} & 143K 	& 0.553 & 0.735 & 0.780 & 0.873 &\textbf{0.878}\\ 
		EastEuropean* 		& 9K 	& 0.488 & 0.517 & 0.466 & 0.575 &\textbf{0.629} & 38K 	& 0.301 & 0.582 & 0.726 & 0.794 &\textbf{0.812}\\ 
		British* 			& 44K 	& 0.611 & 0.760 & 0.789 &\textbf{0.794} & 0.768 & 35K 	& 0.361 & 0.578 & 0.627 & 0.648 &\textbf{0.689}\\ 
		Jewish* 			& 11K 	&\textbf{0.313} & 0.111 & 0.095 & 0.129 & 0.183 & 9K 	& 0.097 & 0.361 & 0.301 &\textbf{0.405} & 0.387\\
		GreaterEastAsian	& 15K 	& 0.637 & 0.626 & 0.642 & 0.690 &\textbf{0.706} & 97K 	& 0.625 & 0.656 & 0.713 &\textbf{0.907} & 0.895\\  
		IndianSubContinent*	& 9K 	& 0.523 & 0.660 & 0.768 &\textbf{0.769} & 0.746 & 26K 	& 0.438 & 0.721 & 0.855 &\textbf{0.912} & 0.903\\ \specialrule{1.5pt}{1pt}{1pt}
		Italian* 			& 14K 	& 0.521 & 0.543 & 0.595 &\textbf{0.634} & 0.613 & 11K 	& 0.233 & 0.453 & 0.665 & 0.713 &\textbf{0.763}\\ 
		Hispanic* 			& 11K 	& 0.403 & \textbf{0.600}& 0.397 & 0.521 & 0.538 & 69K 	& 0.432 & 0.724 & 0.676 & 0.850 &\textbf{0.864}\\ 
		Nordic* 			& 5K 	& 0.400 & 0.587 &\textbf{0.713} & 0.709 & 0.709 & 23K 	& 0.303 & 0.653 & 0.767 &\textbf{0.783}&\textbf{0.783}\\ 
		French* 			& 14K 	& 0.428 & 0.523 & 0.602 & 0.600 &\textbf{0.624} & 27K 	& 0.203 & 0.426 & 0.738 &\textbf{0.769} & 0.750\\ 
		Germanic*			& 5K 	& 0.254 & 0.410 & 0.401 & 0.403 &\textbf{0.412} & 13K 	& 0.140 & 0.431 & 0.582 & 0.629 &\textbf{0.653}\\ 
		Japanese* 			& 8K 	& 0.646 &\textbf{0.724} & 0.456 & 0.547 & 0.695 & 57K 	& 0.674 & 0.788 & 0.434 & 0.928	&\textbf{0.939}\\ 
		EastAsian* 			& 7K 	& 0.499 & 0.455 & 0.609 &\textbf{0.621} & 0.549 & 40K 	& 0.270 & 0.340 & 0.723 &\textbf{0.834} & 0.811\\ \specialrule{1.5pt}{1pt}{1pt}
		Weighted Avg. 			& ---	& 0.492 & 0.607 & 0.619 & 0.648 &\textbf{0.651} & --- 	& 0.364 & 0.580 & 0.642 & 0.790 & \textbf{0.795} \\ \hline
	\end{tabular}
	\caption[test]{F1 scores on a 13-leaf taxonomy. Existing methods: \textit{HMM} \cite{ambekar2009name} and \textit{Ethnea} \cite{torvik2016ethnea};  \textit{Embd} only uses parameters from name embeddings; \ENECWShort is \ENEC with world population as priors. Nationalities on different levels of taxonomy are separated with bold lines. `*' marks leaf nationalities. \textit{Weighted Avg.} is count-weighted average F1 of leaf nationalities.} 
	\label{tab:smallHierarchyPerf}
	\vspace{-0.1in}
\end{table*}

\paragraph{Twitter}
\label{subsubsec:twitterdatacollection}
Although the email data offers the majority of name labels, its imbalanced popularity across the world make some regions inadequate name labels. We noticed that Twitter\footnote{Twitter API: https://dev.twitter.com/rest/public}, as an emerging Web service, has a wider coverage and thus can act as a supplementary source of name labels.

In order to get name labels from interested regions, we \textit{(i)} get list of most popular regional celebrities\footnote{https://www.socialbakers.com/statistics/twitter/profiles/kenya/}; \textit{(ii)} get all followers' Twitter profiles of the celebrities'. Each profile record contains ``name'' and ``location'' fields, though many users leave the latter blank. In summary, we gathered 43M unique Twitter user profiles, within which 9M have non-empty ``location'' field and well-formed names (e.g. two name parts and string length $>$ 1). However, these location tags are not well defined. Among 9M profiles, there are $\sim$1.5M unique locations. Some of them are simply noise, while some offer too much details (e.g. university name without country info.). Therefore, we use Google Map API\footnote{https://developers.google.com/maps/} to query for country names using the 10\% most popular ``locations''. As a result, we have $\sim$6M labeled names for use, supplementary to the labels from email source.

\subsubsection{Name Similarities between countries}
\label{subsubsec:countrySim}
Since our labeled names are collected from Internet, it is important to check its quality. In this subsection, we provide an interesting perspective to validate the high quality of the datasets.

We compute the similarities between countries using the aggregations of names, and check whether they agree with common sense. In fact, we observe that the cultural/spatial closeness between countries are well captured by country name similarities. Take African continent as an example (shown in Fig. \ref{fig:sim_Africa}). On the right-bottom part of the figure, the continent map is divided into 4 major parts based on how close they are culturally and geographically. On the remaining part of the figure, countries with names labels are colored in accordance with the map. It is apparent that countries with same colors are clustered, indicating that nearby countries have similar names. One interesting case is that Angola is connected with Mozambique, even though one is on the west coast of the continent while the other is on the east coast. The reason is that both countries were once colonized by Portuguese, thus many Internet users have Portuguese names. 

We compute the similarities between countries with following steps: (i) aggregate name parts of each country so that countries are represented by name part vectors, where each dimension indicates how many name parts occur in the countries, 
(ii) compute cosine similarity between vectors, i.e. name similarities between countries. Note that in Fig. \ref{fig:sim_Africa}, the thickness of edges indicate the magnitude of similarities. One link is made if either the similarity is larger than 0.5 or it makes sure that each country is linked to at least one most similar countries. Therefore, Ethiopia is linked to Sudan with a very small weight, though it is distinct from other countries.

\begin{table}[!t]
	\centering
	\small
	\begin{tabular}{@{}|c|rcc|rcc|@{}}  \hline
		& \multicolumn{3}{c|}{Wikipedia}  & \multicolumn{3}{c|}{Email/Twitter} \\ \hline
		Nationality & Name\# & \textit{Seer} & \ENECShort &  Name\# & \textit{Seer} & \ENECShort \\ \hline
		Muslim & 7K & 0.560 & \textbf{0.646} & 13K & 0.422 & \textbf{0.688}\\
		EastEuropean & 9K & \textbf{0.739} & 0.596 & 38K & 0.343 & \textbf{0.804}\\
		British & 44K & \textbf{0.852} & 0.843 & 35K & 0.577 & \textbf{0.726}\\
		Indian & 9K & 0.768 & \textbf{0.779} & 26K & 0.639 & \textbf{0.880}\\
		Hispanic & 11K & \textbf{0.605} & 0.558 & 69K & 0.610 & \textbf{0.871}\\
		Germanic & 5K & 0.464 & \textbf{0.487} & 13K & 0.433 & \textbf{0.694}\\
		French & 14K & \textbf{0.676} & 0.650 & 27K & 0.482 & \textbf{0.802}\\
		Italian & 14K & \textbf{0.707} & 0.641 & 11K & 0.329 & \textbf{0.728}\\
		EastAsian & 7K & \textbf{0.824} & 0.635 & 40K & 0.418 & \textbf{0.848}\\
		Japanese & 8K & \textbf{0.875} & 0.550 & 57K & 0.902 & \textbf{0.929}\\\specialrule{1.5pt}{1pt}{1pt}
		Weighted Avg. & --- & \textbf{0.751} & 0.700 & --- &0.571 & \textbf{0.831}\\ \hline
	\end{tabular}
	\caption[test]{F1 scores on 10 nationalities. \textit{EthnicSeer}\cite{treeratpituk2012name} performs slightly better on Wikipedia data but it is an unfair comparison because it is trained on the same dataset. \ENEC performs significantly better on a larger test set from Email/Twitter.} 
	\label{tab:SeerPerf}
	\vspace{-0.25in}
\end{table}

\subsection{Performance Evaluation}
In this Subsection, we will first compare our method with existing systems on smaller nationality taxonomies (one 13-leaf taxonomy and one 10-leaf flat taxonomy \cite{ambekar2009name,torvik2016ethnea,treeratpituk2012name}). Note we use their Web APIs to collect the classification results. Two independent datasets are tested on. The smaller one is from Wikipedia (used in \cite{ambekar2009name,treeratpituk2012name}), the other is from our test set of labeled names. In the end, we will introduce more details about \ENECE's performance on a finer-grained nationality taxonomy.

\subsubsection{On Small Taxonomy}
Ambekar et al. proposed an HMM-based method, which used signals from substrings of names \cite{ambekar2009name} to classify name nationalities. Their taxonomy contains 13 leaf nodes and 18 nodes in total (see \cite{ambekar2009name} for the definition of this taxonomy). In order to compare, all methods need to be on the same taxonomy. HMM is designed on this taxonomy. \ENEC and \textit{Ethnea} are adapted to this because both methods are defined on a finer-grained taxonomy. \textit{EthnicSeer} is compared separately on a flat 10-nationality taxonomy.

Two datasets are available for comparison: \textit{(i)} the labeled names from Wikipedia  (150K in total, the same dataset used to train \textit{HMM} and \textit{EthnicSeer}); \textit{(ii)} we divide Email/Twitter data into training and testing datasets (60\% vs. 40\%). Then we sample 2\% from the test data for use because it is not efficient to get classification results of baselines from their Web APIs (380K). Some small nationalities are given larger sampling ratio to get large enough test samples. 

\begin{table}[t!]
	\centering
	\small
	\begin{tabular}{@{}|crc|crc|@{}} \hline
		Nationality & Name\# & \ENECShort & Nationality & Name\# & \ENECShort\\ \hline
		CelticEnglish* & 3505K & 0.725 & SouthAsian* & 2623K & 0.890\\ 
		Jewish* & 11K & 0.396 & African & 606K & 0.589\\
		Muslim & 1475K & 0.741 & EastAsian & 6157K & 0.920\\
		Greek* & 259K & 0.887 & Hispanic & 6892K & 0.907\\
		Nordic & 195K & 0.731 & European & 5371K & 0.836\\\specialrule{1.5pt}{1pt}{1pt}
		Nubian* & 577K & 0.650 & Japan* & 65K & 0.836\\
		Maghreb* & 47K & 0.148 & Malay & 2596K & 0.863\\
		ArabPeninsula* & 172K & 0.510 & Chinese* & 2901K & 0.928\\
		Turkic & 78K & 0.676 & Portuguese* & 2683K & 0.886\\
		Pakistanis & 179K & 0.511 & Philippines* & 1137K & 0.724\\
		Persian* & 423K & 0.656 & Spanish* & 3072K & 0.851\\
		Finland* & 30K & 0.739 & German* & 1278K & 0.739\\
		Scandinavian & 165K & 0.704 & Baltics* & 12K & 0.408\\
		WestAfrican* & 315K & 0.563 & French* & 2674K & 0.825\\
		SouthAfrican* & 66K & 0.370 & Russian* & 121K & 0.716\\
		EastAfrican* & 225K & 0.574 & EastEurope* & 65K & 0.492\\
		SouthKorea* & 68K & 0.861 & SouthSlavs* & 68K & 0.570\\
		Indochina & 528K & 0.901 & Italian & 1153K & 0.745\\ \specialrule{1.5pt}{1pt}{1pt}
		CentralAsian* & 3K & 0.196 & Cambodia* & 1K & 0.162\\
		Turkey* & 75K & 0.687 & Vietnam* & 502K & 0.913\\
		Bangladesh* & 78K & 0.578 & Thailand* & 18K & 0.592\\
		Pakistan* & 101K & 0.449 & Malaysia* & 242K & 0.480\\
		Denmark* & 49K & 0.662 & Indonesia* & 2354K & 0.870\\
		Sweden* & 74K & 0.607 & Romania* & 329K & 0.663\\
		Norway* & 42K & 0.620 & Italy* & 825K & 0.710\\
		Myanmar* & 7K & 0.607 &  &  & \\ \specialrule{1.5pt}{1pt}{1pt}
		Weighted Avg. & --- & 0.806 &  &  & \\\hline
	\end{tabular}
	\caption{\ENEC performance (F1 scores) on a 39-leaf nationality taxonomy. Nationalities in different levels are separated with bolder lines. `*' marks leaf nationalities. \textit{Weighted Avg.} is count-weighted average F1 of leaf nationalities.} 
	\label{tab:largerHierarchyPerf}
	\vspace{-0.25in}
\end{table}

As shown in Tab. \ref{tab:smallHierarchyPerf}, we compare results of five methods: {\em HMM} \cite{ambekar2009name}, \textit{Ethnea} \cite{torvik2016ethnea}, \textit{Embd}, \ENEC and \ENECWE. \textit{Embd} only use parameters estimated from name embeddings. \ENECW uses the world population as priors. \ENEC and \ENECW performs best on most classes for both datasets. On Wikipedia data, our methods achieves best performances on 15 (out of 18) classes. Some classes get +10\% F1 boost, including Indian, Nordic and EastAsian. On Email/Twitter data, the improvement is more significant. \ENEC outperforms the rest on all classes. Some classes get performance increase by +30\%, including Muslim, Africans, etc. Note that \textit{Embd} also achieves considerable high performance, indicating that name embedding is capturing nationality signals well.

\begin{figure*}[t!]
	\centering
	\begin{overpic}[width=0.28\textwidth]{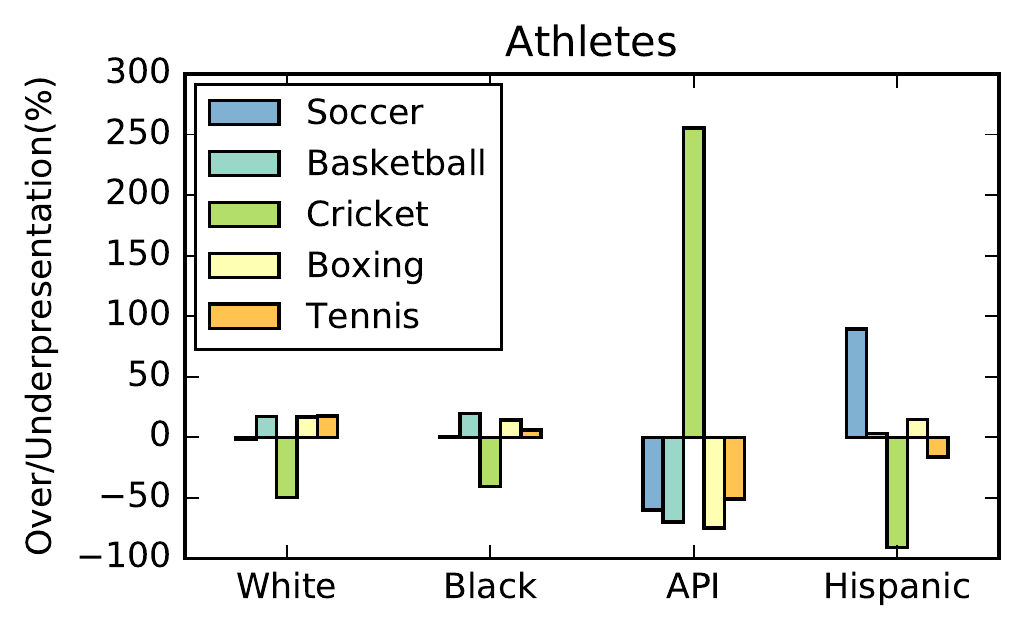}
		\put (42,56) {(a)}
	\end{overpic}
	\begin{overpic}[width=0.23\textwidth]{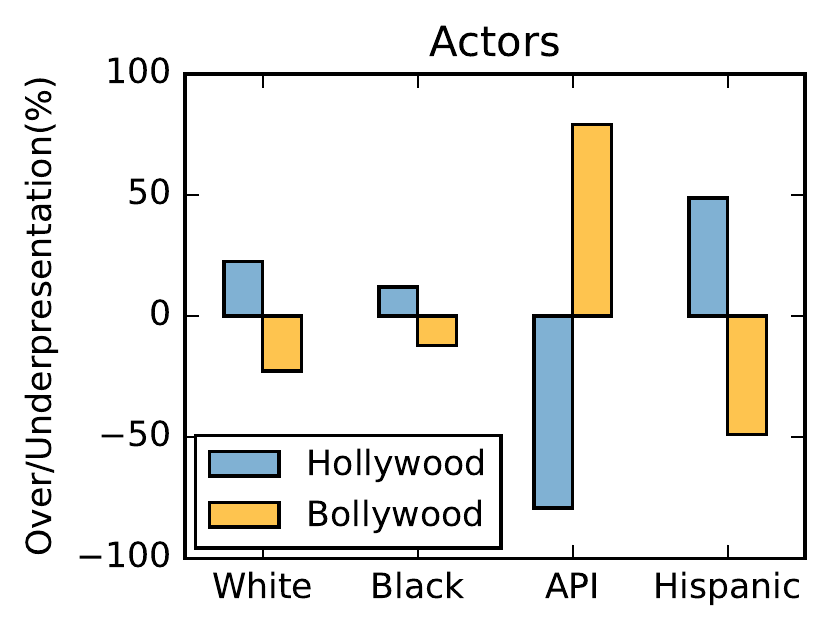}
	\put (42,69) {(b)}
	\end{overpic}
	\begin{overpic}[width=0.23\textwidth]{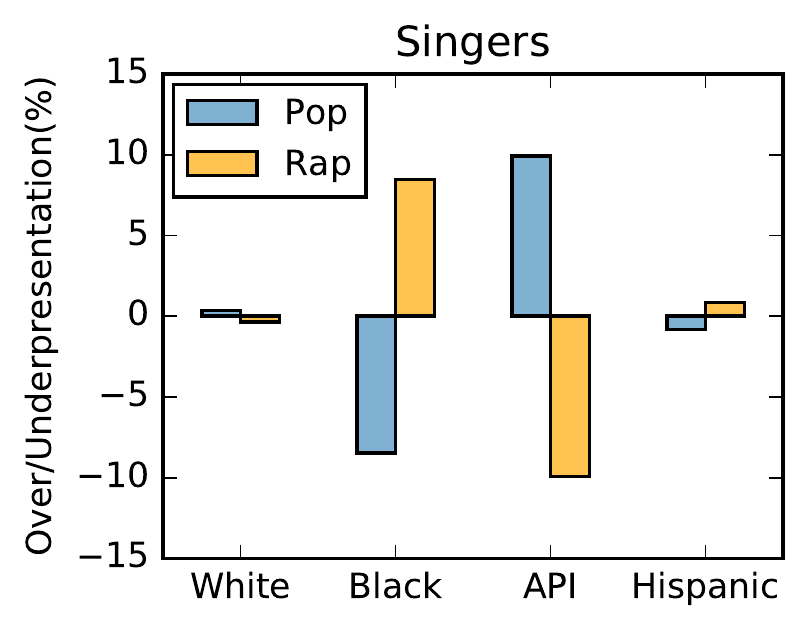}
	\put (40,71) {(c)}
	\end{overpic}
	\begin{overpic}[width=0.23\textwidth]{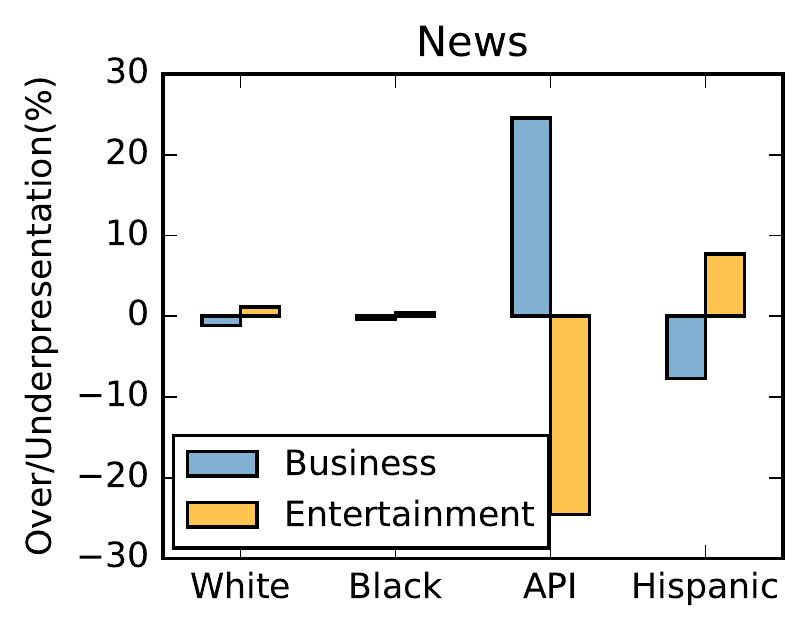}
	\put (42,71) {(d)}
	\end{overpic}
	\caption{Ethnicity Over/underrepresentation of U.S. Twitter users' interest on different topics: (a) Cricket is almost exclusively followed by Indians while soccer is more popular among Hispanics. (b) U.S. actors enjoy a more diverse popularity than Indian actors. (c) African-Americans like rap more than pop. (d) Asians follow business news more than entertainment.}
	\label{fig:app4topics}
\end{figure*}

\textit{EthnicSeer} is defined on a 10-leaf flat taxonomy. For comparison purpose, we removed the labeled names from African, Jewish and Nordic from both datasets. We also shrink \ENECE's 39-leaf taxonomy to fit this small one. The weighted average F1 score shows \textit{EthnicSeer} performs slightly better on Wikipedia but it is the same dataset that 
\textit{EthnicSeer} is trained on. In contrast,  \ENEC performs significantly better on Email/Twitter testing set.

\subsubsection{On Large Taxonomy}
Tab. \ref{tab:largerHierarchyPerf} shows \ENEC F1 scores on the large nationality taxonomy. Note we randomly split the Email/Twitter data into training and testing sets (60\% vs. 40\%) for 3 times. All reported performances of our methods (i.e. \textit{Embd}, \ENEC and \ENECWE) are average F1 of 3 runs. The standard deviations are all below 0.005. As we can see from Tab. \ref{tab:largerHierarchyPerf}, \ENEC performs well on most nationalities. For some less developed countries with few Internet users, including Central Asian countries and Maghreb countries, we have limited number of name labels and contact lists. Thus the performances on these nationalities are limited. To the best of our knowledge, our work is the first effort trying to classify names belonging to these regions.

\begin{figure}[t!]
	\centering
	\begin{overpic}[width=0.46\textwidth]{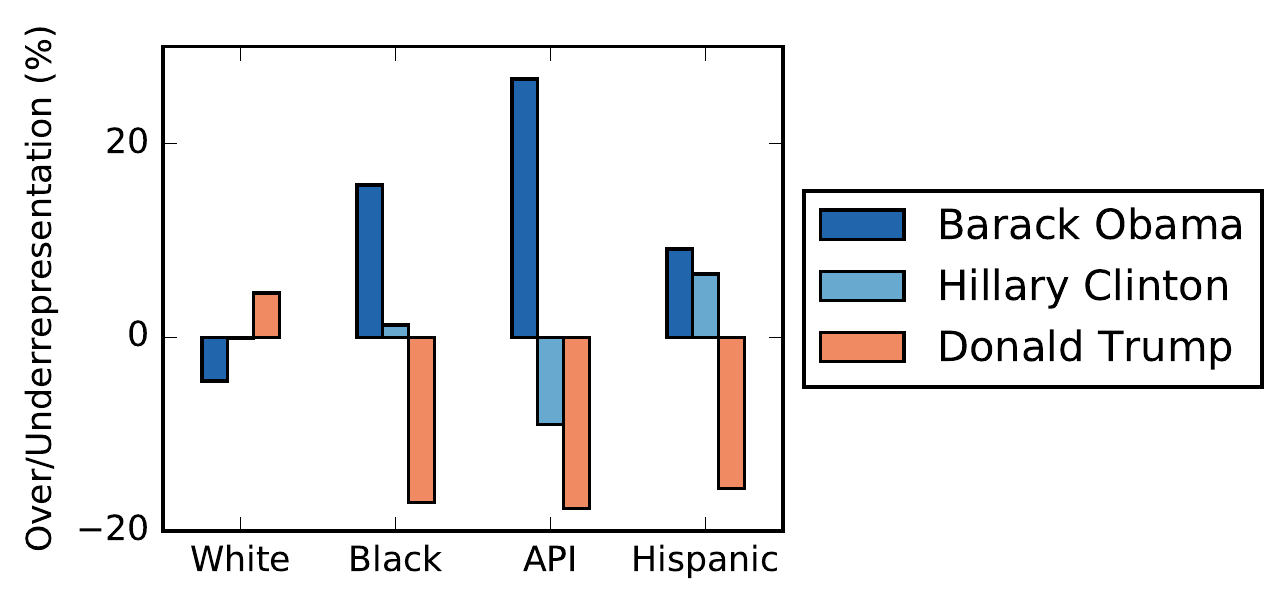}
	\end{overpic}
	\caption{Ethnicity Over/underrepresentation of Barack Obama, Hillary Clinton and Donald Trump's U.S. Twitter followers. White followers are overrepresented for Trump, Obama and Clinton have more followers among minorities.}
	\label{fig:appPresidents}
\end{figure}

\section{Ethnicity Classification}
\label{sec:ethnicity}

As we have mentioned in Sec. \ref{sec:embedding}, U.S. Census Bureau defined 6 race/ethnicity: White, Black, API, Hispanic, AIAN and 2PRACE. In order to build classifier for these ethnicities, we need labeled names for these ethnicities to estimate parameters. Fortunately, U.S. Census Bureau published ethnicity distribution for popular last names. We can estimate first names' ethnicity distribution by connecting census labels with email names from the U.S.

More formally, let $V_{us}^L$ be the set of popular last names from Census Bureau, so we have ground truth, $P_{us}(E|v_l), \forall v_l \in V_{us}^L$, where $E$ denote ethnicity. We can estimate the posteriors of first names with Eq. \ref{equ:estimateFirst}.

\begin{equation}
\label{equ:estimateFirst}
P_{us}(E|v_f) = \frac{1}{|S(v_f)|}\sum\limits_{v_l \in S(v_f)}{P(E|v_l)}
\end{equation}

\noindent where $S(v_f)$ is the list of last names and $(v_f, v_l)$ is a full name from U.S. email data. Note that some of the last names paired with $v_f$ may not have a ground truth label (i.e. $v_l \notin V_{us}^L$). To make reliable estimation, we only keep first names that at least half of the paired last names with a ground truth label. Therefore, we form a set of first names ($V_{us}^F$) with estimated ethnicity distributions. We denote $V_{us} = V_{us}^F \cup V_{us}^L$. We can get $P_{us}(v_f|E)$ and $P_{us}(v_l|E)$ by applying Bayes Rules.

For now, $V_{us}$ can handle names with popular first/last names. For rare names, we can make use of the Email/Twitter name labels. 118 countries are assigned to the six ethnicities based on their definitions. For example, we make names from European countries as White while names from Asian as API. Therefore, we can follow similar steps as Algorithm \ref{alg:hierarchical}. The difference is we will first check whether a name part is from $V_{us}$. If yes, we will use $P_{us}(v_i|E)$ to compute $P(E|v_f,v_l)$ because they are estimated from ground truth with high confidence. Otherwise, we will then check whether they are in $V_{tr}$ or $V_{em}$ as in Algorithm \ref{alg:hierarchical} and follow the remaining steps.

\section{Social Media Analysis}
\label{sec:app}

Nationality and ethnicity classification have broad application in 
sociological research and media analysis.
Here we present some interesting observations, when we apply our classifiers
to the followers of Twitter celebrities.

To collect data, we identified the 100 most followed celebrities in each
of six categories: actors, singers, news, atheletes, governments and politicians;
all of whom have from 1M to 100M followers.
For each celebrity, we selected 50,000 random followers,
and filtered out accounts with irregular names using the same method as
discussed in \ref{subsubsec:twitterdatacollection}).
We then apply \ENEC and \prism to the remaining followers. 
\begin{figure}[t!]
	\centering
	\includegraphics[width=0.4\textwidth]{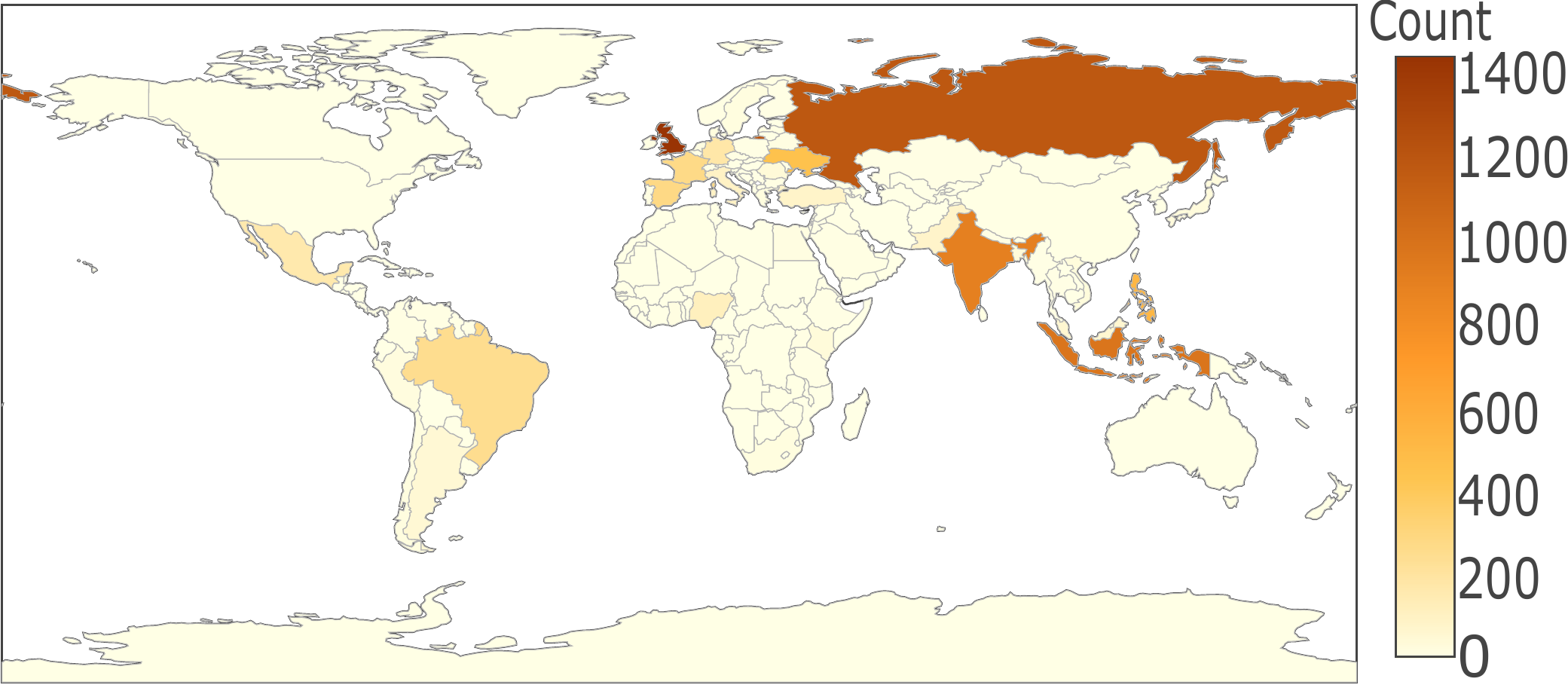}
	\caption{An Indonesian politician has 50\% followers with British, Russian or Indian names while only 13\% are Indonesians. Besides, his Twitter profile is also suspicious: 23K Tweets but only 1 following. His tweets are written in Indonesian, which most followers can not understand.}
	\label{fig:appSuspCele}
	\vspace{-0.1in}
\end{figure}%

Our primary observations here include:

\begin{itemize}
	\item
	{\em Ethnicity and the 2016 U.S. Presidential Election} --
	There has been considerable concern that the recent election 
	exacerbated tensions between ethnic groups in the United States.
	Indeed, our analysis of U.S.-based followers of the primary figures in the race
	(Obama, Clinton, and Trump) show stark differences in composition.
	Fig. \ref{fig:appPresidents} shows that whites are substantially
	overrepresented among Trump's followers, while Clinton and
	Obama have disproportionately more followers among minorities.

	\item
	{\em Interests and Ethnicity} --
	Fig. \ref{fig:app4topics} similarly breaks down the followers of major
	celebrities in sports, entertainment, and news categories.
	The followers of cricket and Bollywood stars are overwhelmingly Indian,
	while Hispanics disproportionally favor soccer and boxing.
	
	\item
	{\em Anomaly Detection through Nationality Analysis} --
	We were surprised to learn that an Indonesian politician named Jeffrie Geovanie
	was one of the most heavily followed figures on Twitter,
	because he has only 45K Google search results about him, mostly in Indonesian.
	Yet our name analysis of his followers shows that only 13\% are Indonesian,
	with over 50\% of the followers of British, Russian, or Indian nationality (Fig. 9).
	This is quite peculiar given that Indonesian is the primary language
	of his Twitter stream.
\end{itemize}

\section{Conclusion}
\label{sec:conclusion}
We demonstrate that homophily patterns in communications can be
exploited to learn name embeddings, that capture interesting properties of
gender, nationality and ethnicity. Further we use these embeddings to build state-of-the-art name nationality and ethnicity classifiers. Through extensive experiments, we show that \ENEC substantially outperforms exiting methods on two independent datasets. Finally, we apply our classification to the Twitter celebrities' followers, with interesting results.

We believe that \ENEC will become an important tool for biomedical and
sociological research. Future work revolves around applying name embeddings to other classification tasks, such as those arise in demographics, security and social media analysis.

\bibliographystyle{ACM-Reference-Format}
\bibliography{sigproc} 

\end{document}